\documentclass[twocolumn,showpacs,preprintnumbers,amsmath,amssymb]{revtex4}
\usepackage{graphicx}
\usepackage{dcolumn}
\usepackage{bm}

\newcommand{\bef}{\begin{figure}}
\newcommand{\eef}{\end{figure}}

\newcommand{\be}{\begin{equation}}
\newcommand{\ee}{\end{equation}}
\newcommand{\bea}{\begin{eqnarray}}
\newcommand{\eea}{\end{eqnarray}}
\usepackage{xspace}     
\usepackage{color}

\newcommand{\sqrts}[1]{\mbox{$\sqrt{s_{_{NN}}}$}~=~#1~GeV\xspace} 

\begin{document}

\title{Selected Experimental Results from Heavy Ion Collisions at LHC}

\author{Ranbir Singh$^1$, Lokesh Kumar$^{2,3}$, Pawan Kumar Netrakanti$^4$ and Bedangadas Mohanty$^3$}
\affiliation{
$^1$Physics Department, University of Jammu, Jammu 180001, India, \\
$^2$Kent State University, Kent, Ohio 44242, USA, \\
$^3$School of Physical Sciences, National Institute of Science Education and Research, Bhubaneswar 751005, India, and \\
$^4$Nuclear Physics Division, Bhabha Atomic Research Centre, Mumbai 400 085, India
}

\date{\today}
\begin{abstract}
We review a subset of experimental results from the heavy-ion collisions at the Large Hadron Collider 
(LHC) facility at CERN. 
Excellent consistency is observed
across all the experiments at the LHC (at center of mass energy
$\sqrt{s_{\mathrm {NN}}} = 2.76$ TeV) 
for the measurements such as charged
particle multiplicity density, azimuthal anisotropy
coefficients and nuclear modification factor of charged hadrons.
Comparison to similar measurements from the Relativistic Heavy
Ion Collider (RHIC) at lower energy ($\sqrt{s_{\mathrm {NN}}} =$ 200 GeV) suggests that system formed
at LHC has a higher energy density, larger system size, and lives
for a longer time. These measurements are compared to model calculations
to obtain physical insights on the properties of matter created at the
RHIC and LHC.

\end{abstract}
\maketitle

\section{Introduction}
\label{sec:introduction}

The main goal of the high energy heavy-ion collisions is to study the phase structure of the Quantum Chromodynamic
(QCD) phase diagram\cite{Mohanty:2011ki,Mohanty:2009vb,Fukushima:2010bq}. 
One of the most interesting aspects of these collisions is the possibility of forming a phase of 
de-confined quarks and gluons, a system that is believed to have existed in a few microsecond old Universe. First 
principle QCD calculations suggest that it is possible to have such a state of matter if the temperatures attained
can be of the order of the QCD scale ($\sim$ 200 MeV)~\cite{Gupta:2011wh,Bazavov:2011nk,Borsanyi:2010bp}. 
In laboratory, such temperatures could be attained by colliding
heavy-ions at relativistic energies. 
Furthermore, in very high energy collisions of heavy-ions at the LHC
  and RHIC, the lifetime of  the deconfined phase may be long enough to allow for the detailed
  study of the fundamental constituents (quarks and gluons) of the
  visible matter.

The results from heavy-ion collisions at RHIC have clearly demonstrated the formation of a de-confined system
of quarks and gluons in Au+Au collisions at $\sqrt{s_{\mathrm {NN}}} = 200$ GeV~\cite{Arsene:2004fa,Back:2004je,Adams:2005dq,Adcox:2004mh,Gyulassy:2004zy}.  The produced system exhibits
copious production of strange hadrons, shows substantial collectivity developed in the partonic phase, exhibits
suppression in high transverse momentum ($p_{T}$) hadron production relative to $p$+$p$ collisions and
small fluidity as reflected by a small value of viscosity to entropy density ratio ($\eta/s$).
A factor of 14 increase in $\sqrt{s_{\rm{NN}}}$ for Pb+Pb collisions at LHC is
expected to unravel the temperature dependence of various observables, as
well as to extend the kinematic reach in rapidity and $p_T$ of previous
measurements at RHIC. On the other hand, the beam energy scan program at
RHIC is expected to provide additional details of the QCD phase diagram
not accessible at the LHC~\cite{Mohanty:2011nm}.

In this review paper, we discuss a subset of results that have come out from LHC Pb+Pb collisions 
at $\sqrt{s_{\mathrm {NN}}} = 2.76$ TeV. We have divided the discussion into three sections. In the second
section we discuss the consistency of various measurements among the three LHC experiments that have 
heavy-ion programs: ALICE, ATLAS and CMS
Subsection IIA discusses the results on the charged
  particle multiplicity, IIB on azimuthal anisotropy, and IIC on the
  nuclear modification factor.

In the third section, we make a comparative study between 
similar observables measured at lower energy collisions at RHIC to those from LHC. In doing this, 
we highlight the additional information that heavy-ion collisions at LHC bring compared to RHIC. 
In subsection IIIA, we discuss the bulk properties at freeze-out that
include results on multiplicity, average transverse mass and Bjorken
energy density, volume and decoupling time, kinetic freeze-out
temperature and average flow velocity, and fluctuations.
Subsection IIIB is devoted on the results on azimuthal anisotropy where we
discuss the energy dependence of $p_T$ integrated $v_2$, dependence of
various azimuthal anisotropy coefficients on $p_T$, and flow
fluctuations.
In subsection IIIC, we discuss results for nuclear modification factor.

In the fourth section, we present a comparison of various model calculations to the corresponding 
measurements at LHC. 
We concentrate mainly on the results for charged particle multiplicity
density and $K/\pi$ ratio in subsection IVA, azimuthal anisotropy in subsection IVB,
and nuclear modification factor in subsection IVC. 

Finally, we summarize our observations in the last section of the article.
\section{Consistency of results among LHC experiments}
\label{sec:clhc}
\subsection{Charged particle multiplicity}
  \begin{figure}
\includegraphics[scale=0.4]{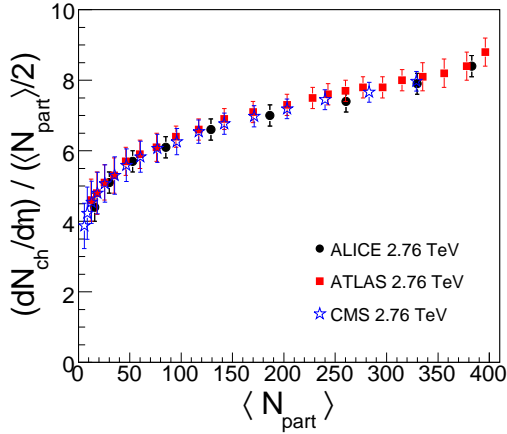}
    \caption{\label{fig:dndetalhc}
      (Color online) Average charged particle multiplicity per unit pseudorapidity ($dN_{\rm ch}/d\eta$) 
at midrapidity per participating nucleon ($\langle N_{\rm {part}}
\rangle$) pair plotted as a function of $\langle N_{\rm {part}} \rangle$ for 
Pb+Pb collisions at $\sqrt{s_{\mathrm {NN}}}$ = 2.76 TeV. The measurements are shown from ALICE~\cite{Aamodt:2010cz}, 
CMS~\cite{Chatrchyan:2011pb} and ATLAS~\cite{ATLAS:2011ag} experiments. }
  \end{figure}
One of the first measurements to come out of the heavy-ion collision program at LHC is the
charged particle multiplicity per unit pseudorapidity in Pb+Pb collisions at $\sqrt{s_{\mathrm {NN}}}$ = 2.76 TeV.
Figure~\ref{fig:dndetalhc} shows the centrality (reflected by  the number of participating nucleons, 
$N_{\rm {part}}$, obtained from a Glauber model calculation~\cite{Miller:2007ri}) dependence of $dN_{\rm ch}/d\eta$
at midrapidity for Pb+Pb collisions at $\sqrt{s_{\mathrm {NN}}}$ = 2.76 TeV from ALICE~\cite{Aamodt:2010cz}, CMS~\cite{Chatrchyan:2011pb} and ATLAS~\cite{ATLAS:2011ag}  experiments. The error bars reflect statistical uncertainties. 
The ATLAS measurements of  $dN_{\rm ch}/d\eta |_{\eta =0}$ are 
obtained over $|\eta|$ $<$ 0.5 using a minimum bias trigger with a central solenoid magnet off data set. The charged 
particles are reconstructed using two different algorithms using the information from pixel detectors covering $|\eta| <$ 2.0. 
The $N_{\rm {part}}$ values are obtained by comparing the summed transverse energy in the forward calorimeter over 
a pseudorapidity range 3.2 $<$ $|\eta|$ $<$ 4.9 to a Glauber model simulation.  
The CMS results for $dN_{\rm ch}/d\eta |_{\eta =0}$ are from the barrel section of the  
pixel tracker covering $|\eta| <$ 2.5. The minimum bias trigger data set was in the magnetic field off configuration so as to 
improve the acceptance of low $p_{T}$ particles. The centrality determination as in the case of ATLAS experiment are
done using information from hadron forward calorimeter (2.9 $<$ $|\eta|$ $<$ 5.2) and Glauber model simulations.
The ALICE measurement uses a minimum bias data set from the silicon pixel detector ($|\eta| <$ 2.0). The
centrality selection is carried out using signals from VZERO detectors (2 arrays of 32 scintillators tiles) 
covering the region 2.8 $<$ $\eta$ $<$ 5.1 and  -3.7 $<$ $\eta$ $<$ -1.7, along with the corresponding Glauber 
modeling of the data. 

In-spite of difference in operating conditions and measurement techniques, the $dN_{\rm ch}/d\eta$
versus $N_{\rm {part}}$ results for Pb+Pb collisions at $\sqrt{s_{\mathrm {NN}}}$ = 2.76 TeV shows a remarkable 
consistency across the three experiments. 
The results show that the charged particle multiplicity per unit pseudorapidity per nucleon pair
increases from peripheral to central collisions.
This gradual increase in $dN_{\rm ch}/d\eta$ per participating
nucleon pair indicates that in central head-on collisions, where the number of participating nucleons are more, the charged particle
production is different than as compared to that in peripheral collisions.

\subsection{Azimuthal anisotropy}

\begin{widetext}
Azimuthal anisotropy has been studied in great details in heavy-ion
collision experiments. It can provide information about initial stages
of heavy-ion collisions.
Figure~\ref{fig:vnlhc} (top panels)
shows the azimuthal anisotropy of produced charged particles ($v_{n}=\langle\cos(n(\phi-\Psi_{n}))\rangle$) 
as a function of $p_{T}$ for 30--40\% Pb+Pb collisions at
$\sqrt{s_{\mathrm {NN}}}$ = 2.76 TeV  from the three different experiments: ATLAS, ALICE and CMS.
Here, $\phi$ is the azimuthal angle of the produced particles and $\Psi_{n}$ is the $n^{th}$ order 
reaction plane angle measured in the experiments.  
Left panel in the figure corresponds to $v_{2}$, middle panel to $v_{3}$, and right
panel to $v_{4}$, respectively.
Bottom panels show the ratio of the experimental data to a polynomial fit to the ALICE data.

In the CMS experiment~\cite{Chatrchyan:2012xq}, the $v_{2}$ measurements use the information from the silicon tracker 
in the region $|\eta| <$ 2.5
with a track momentum resolution of 1\% at $p_{T}$ = 100 GeV/$c$ and kept within a magnetic field of 3.8 Tesla. 
The event plane angle ($\Psi_{2}$) is obtained using the 
information on the energy deposited in the hadron forward calorimeter. A minimum $\eta$ gap of 3 units is kept between the particles
used for obtaining $\Psi_{2}$ and $v_{2}$. This ensures suppression of non-flow correlations which could arise for example from dijets. 
The event plane resolution obtained using three sub events technique varies from 0.55 to 0.84, depending on the collision centrality. 
The ATLAS experiment~\cite{ATLAS:2011ah} measured $v_{n}$ using the inner detectors in the $|\eta| <$ 2.5, 
kept inside a 2 Tesla field of superconducting solenoid magnet. The event planes are obtained using forward calorimeter information,
with resolution varying from 0.2 to 0.85 depending on collision centrality.  The ALICE experiment~\cite{Aamodt:2010pa} measured $v_{n}$
using charged tracks reconstructed from the Time Projection Chamber ($|\eta| <$ 0.8), the event plane was obtained using
information from VZERO detectors kept at a large rapidity gap from the TPC. The momentum resolution of the tracks are better
than 5\%. 

A very nice agreement for $v_{2}$, $v_{3}$, and $v_{4}$ versus $p_{T}$
is found between all the experiments to a level of within 10\% for 
most of the $p_{T}$ range presented. 
The results show an increase of $v_2$, $v_3$, and $v_4$ values with $p_T$ for the low $p_T$ and then decreases for
$p_T$ above $\sim$3 GeV/$c$. The hydrodynamical evolution of the
system affects most of the low $p_T$ particles and hence the increasing $v_n$ at low $p_T$.

  \begin{figure}
\includegraphics[scale=0.65]{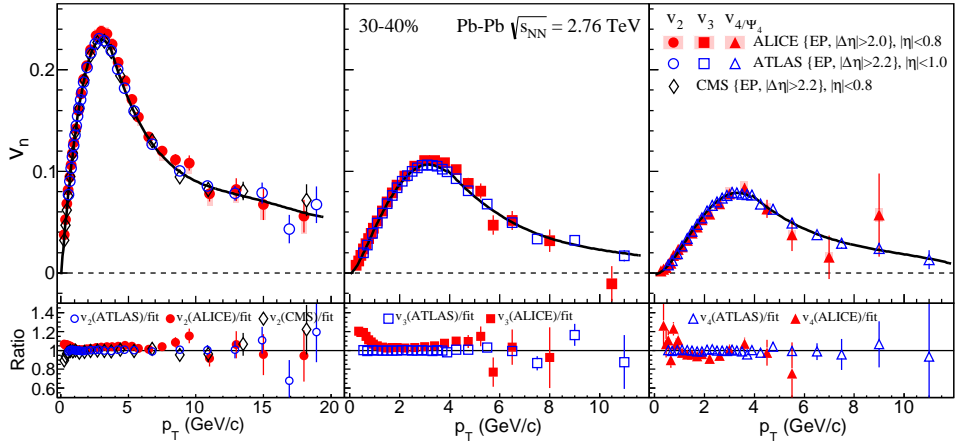}
    \caption{\label{fig:vnlhc}
      (Color online) $v_{\rm n}$ versus $p_{T}$ at midrapidity for 30-40\% Pb+Pb collisions at 
 $\sqrt{s_{\mathrm {NN}}}$ = 2.76 TeV. The results are shown from different LHC experiments: CMS~\cite{Chatrchyan:2012xq}, ATLAS~\cite{ATLAS:2011ah} and ALICE~\cite{Aamodt:2010pa}. The bottom panels shows the ratio of the experimental data to a polynomial fit to the
ALICE data.}
  \end{figure}

\end{widetext}

\subsection{Nuclear modification factor}

\begin{figure}
\includegraphics[scale=0.4]{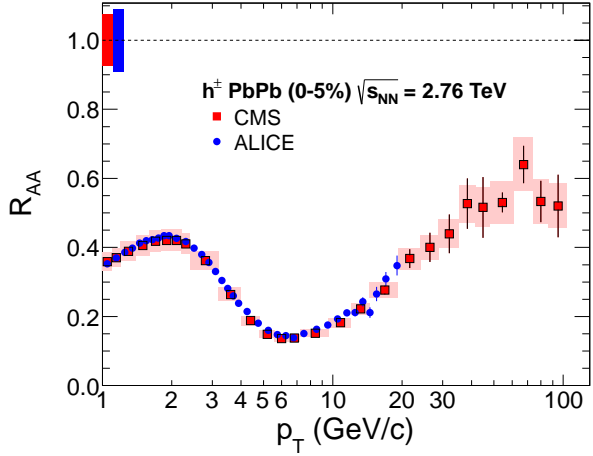}
\caption{\label{fig:raalhc}
     (Color online) Nuclear modification factor $R_{\rm {AA}}$ of charged hadrons measured by ALICE~\cite{Aamodt:2010jd} and CMS~\cite{CMS:2012aa} 
experiments at midrapidity for 0-5\% most central Pb-Pb collisions at $\sqrt{s_{\mathrm {NN}}}$ = 2.76 TeV. 
The boxes around the data denote $p_{T}$-dependent systematic uncertainties. 
The systematic uncertainties on the normalization are shown as boxes at $R_{\rm {AA}}$ = 1.}
\end{figure}

One of the established signature of the QGP at top
 RHIC energy is the suppression of high transverse momentum ($p_T$)
  particles in heavy-ion collisions compared to corresponding data
  from the binary collisions scaled $p+p$ collisions. It has been
  interpreted in terms of energy loss of partons in QGP. This
  phenomenon is referred to as the jet quenching in a dense partonic
  matter. The corresponding measurement is called the nuclear
  modification factor ($R_{AA}$). 

Figure~\ref{fig:raalhc} shows the nuclear modification factor for inclusive charged hadrons measured at midrapidity 
in LHC experiments for Pb+Pb collisions at $\sqrt{s_{\mathrm {NN}}}$~=~2.76 TeV. The nuclear modification
factor is defined as $R_{AA}=\frac{dN_{AA}/d\eta d^2 p_{T}} {T_{AB} d\sigma_{NN}/d\eta d^2 p_{T}}$,
here the overlap integral $T_{AB} = N_{binary}/\sigma_{inelastic}^{pp}$ with $N_{binary}$ being the
number of binary collisions commonly estimated from Glauber model calculation and $d\sigma_{NN}/d\eta d^2 p_{T}$
is the cross section of charged hadron production in $p$+$p$ collisions at $\sqrt{s}$ = 2.76 TeV.

The ALICE experiment~\cite{Aamodt:2010jd} uses the Inner Tracking System (ITS) and the Time Projection
Chamber (TPC) for vertex finding and tracking in a minimum bias data set. 
The CMS experiment~\cite{CMS:2012aa} reconstructs charged particles based
on hits in the silicon pixel and strip detectors. In order to extend the statistical reach of the
$p_{T}$ spectra in the highly pre-scaled minimum bias data recorded in 2011, it uses unprescaled single-jet triggers. 
Both experiments take the value of $\sigma_{inelastic}^{pp}$ = 64 $\pm$ 5 mb. 
The result shows that the charged particle production at high $p_{T}$ in LHC is suppressed in heavy-ion collisions relative to
nucleon-nucleon collisions. 
The suppression value reaches to a minimum at $p_{T}$ 6-7 GeV/$c$ and then gradually increases to attain
an almost constant value at $\sim$ 40 GeV/$c$. This can be understood in terms of energy loss mechanism differences in intermediate and
higher $p_{T}$ regions. The rise in the $R_{AA}$ above $p_{T}$ 6-7 GeV/$c$ may imply the dominance of the constant fractional
energy loss which is the consequence of flattening of the unquenched nucleon-nucleon spectrum.
An excellent agreement for $R_{\rm {AA}}$ versus $p_{T}$ for charged hadrons in 0-5\% central Pb+Pb collisions 
at $\sqrt{s_{\mathrm {NN}}}$ = 2.76 TeV is observed between the two experiments.

Having discussed the consistency of these first measurements in Pb+Pb collisions among different experiments,
the major detectors used, acceptances and ways to determine centrality and event plane, we now discuss the
comparison between measurements at RHIC and LHC heavy-ion collisions.

\section{Comparison of LHC and RHIC results}
\label{sec:results}

In the first subsection, we discuss the energy dependence of basic measurements made in heavy-ion collisions.
These include $dN_{\rm {ch}}/d\eta$, $\langle m_{T} \rangle$ ($m_{T} =
\sqrt{p_{T}^{2} + m^{2}}$, here $m$ represents mass of hadron), 
Bjorken energy density ($\epsilon_{Bj}$), life time of the hadronic phase ($\tau_{f}$), system 
volume at the freeze-out, kinetic and chemical freeze-out conditions and finally, the fluctuations in net-charge
distributions.  
In the next subsection, we discuss about the energy dependence of
$p_{T}$ integrated $v_{2}$, $v_{n}$ versus $p_{T}$ and flow fluctuations 
at RHIC and LHC.
In the final subsection, we compare the nuclear modification factor for hadrons produced 
in heavy-ion collisions at RHIC and LHC.

\subsection{Bulk properties at freeze-out}

\subsubsection{Multiplicity}

 %
  \begin{figure}[htbp]
\includegraphics[scale=0.4]{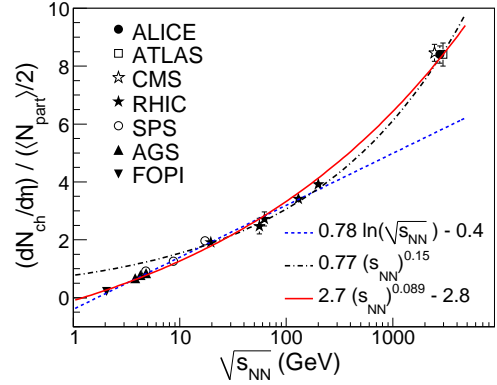}
\includegraphics[scale=0.4]{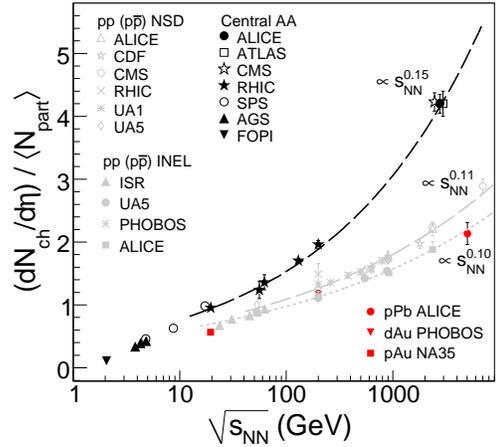}
    \caption{\label{fig:dnchetacmen}
      (Color online) Top panel: $dN_{ch}/d\eta$ per participating nucleon pair at midrapidity in central heavy ion collisions 
as a function of $\sqrt{s_{\rm {NN}}}$.  Bottom panel: Comparison of $dN_{ch}/d\eta$ per participating nucleon at
midrapidity in central heavy-ion collisions~\cite{Back:2003hx,Abreu:2002fx,Adler:2001yq,Bearden:2001xw,Bearden:2001qq,Adcox:2000sp,Back:2000gw,Back:2002wb,Aamodt:2010pb,ATLAS:2011ag,Chatrchyan:2011pb} to corresponding results from $p$+$p$($\bar{p}$)~\cite{ua1,ua5,Abelev:2008ab,Abe:1989td,Aamodt:2010pp,Khachatryan:2010xs,isr,Aamodt:2010ft,Alver:2010ck}  and $p(d)$+A collisions~\cite{ALICE:2012xs,Alber:1997sn,Back:2003hx}. }
  \end{figure}

Figure~\ref{fig:dnchetacmen} (top panel) shows the charged particle multiplicity density at midrapidity ($dN_{ch}/d\eta$) per 
participating nucleon pair produced in central heavy-ion collisions versus  $\sqrt{s_{\rm {NN}}}$. We observe that
the charged particle production increases by a factor 2 as the energy increases from RHIC to LHC. The energy
dependence seems to rule out a logarithmic dependence of particle production with  $\sqrt{s_{\rm {NN}}}$ and 
supports a power law type of dependence on $\sqrt{s_{\rm {NN}}}$. 
The red solid curve seems to describe the full energy
range.
More detailed discussions on the energy dependence of these measurements 
can be found in the Ref.~\cite{mishra}.

Figure~\ref{fig:dnchetacmen} (bottom panel) 
shows the excess of $dN_{ch}/d\eta$/$\langle N_{part} \rangle$ in A+A collisions~\cite{Back:2003hx,Abreu:2002fx,Adler:2001yq,Bearden:2001xw,Bearden:2001qq,Adcox:2000sp,Back:2000gw,Back:2002wb,Aamodt:2010pb,ATLAS:2011ag,Chatrchyan:2011pb} 
 over corresponding yields in $p$+$p$($\bar{p}$)~\cite{ua1,ua5,Abelev:2008ab,Abe:1989td,Aamodt:2010pp,Khachatryan:2010xs,isr,Aamodt:2010ft,Alver:2010ck}
 and $p(d)$+A collisions\cite{ALICE:2012xs,Alber:1997sn,Back:2003hx}. 
This observation also seen at RHIC persists at LHC 
but is proportionately larger at the higher energy collisions at the LHC. A power law fit to the $p$+$p$ collision 
charged particle multiplicity density leads to a dependence $\sim$ $s^{0.11}$, while those for A+A collisions 
goes as $\sim$ $s^{0.15}$. There is no scaling observed in the charged particle multiplicity density
per participating nucleon, when compared  between elementary collisions like $p$+$p$ and heavy-ion 
collisions. This is a clear indication that A+A collisions at RHIC  and LHC are not a simple 
superposition of several $p$+$p$ collisions, whereas the $p$+A collisions scale with the 
$p$+$p$ collisions.

\subsubsection{Average transverse mass and Bjorken energy density}

 %
  \begin{figure}[htbp]
\includegraphics[scale=0.4]{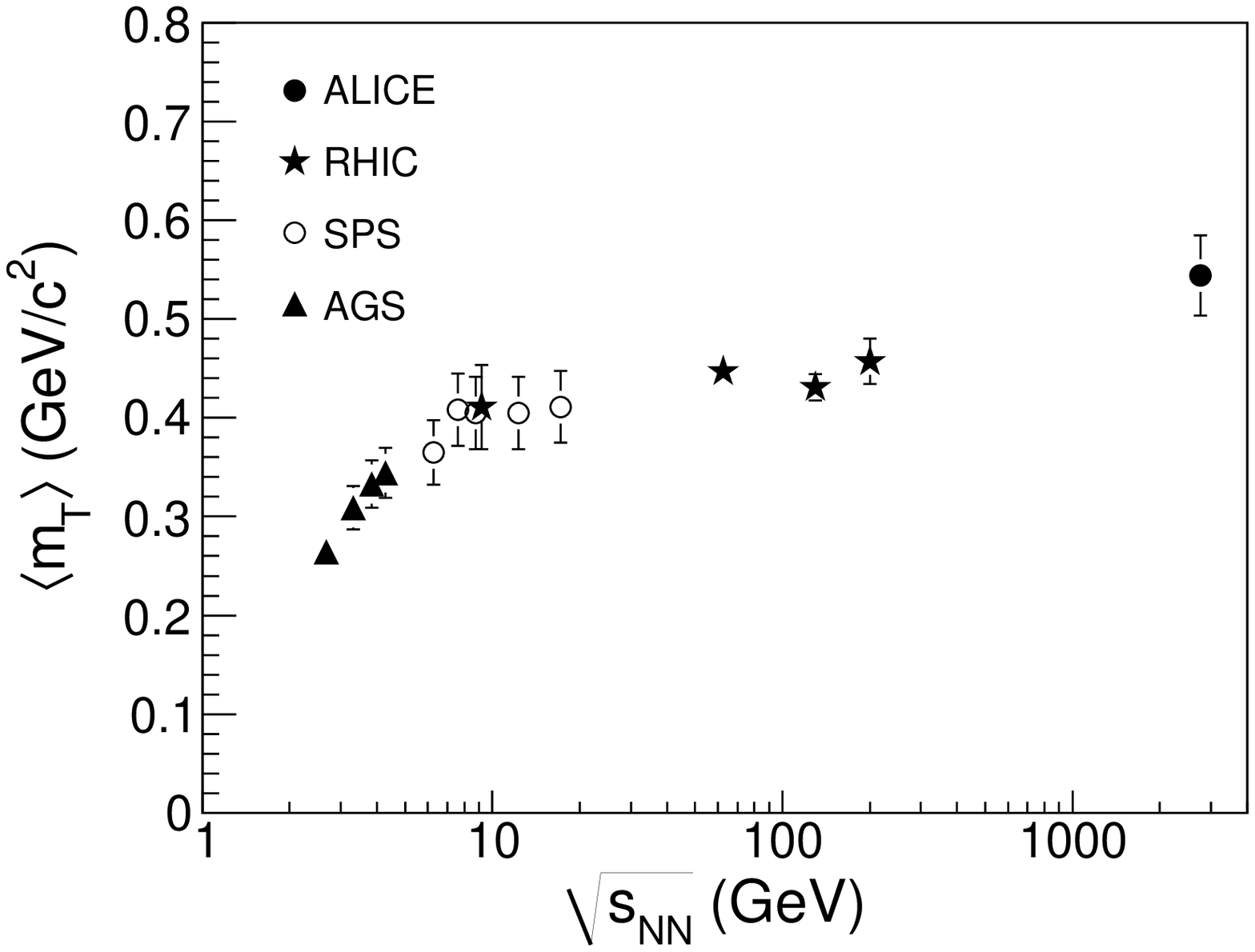}
\includegraphics[scale=0.4]{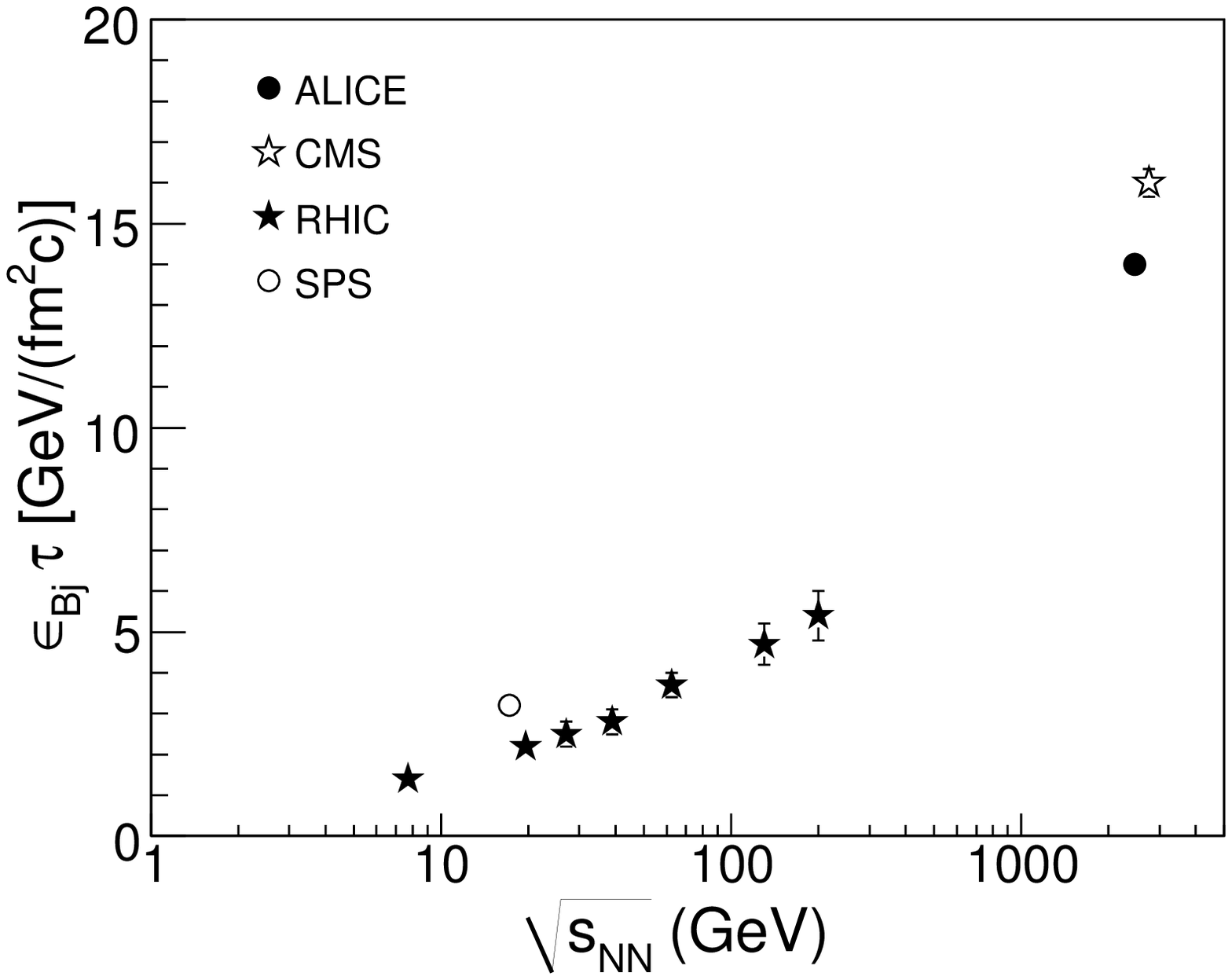}
    \caption{\label{fig:den}
      Top panel: $\langle m_{T} \rangle$ for charged pions in central heavy-ion collisions at
midrapidity for AGS~\cite{Ahle:1999uy,Barrette:1999ry}, SPS~\cite{Afanasiev:2002mx,Alt:2007aa}, RHIC~\cite{Abelev:2008ab,Abelev:2009bw}
and LHC~\cite{Abelev:2012wca}  energies. The errors shown are the quadrature sum of statistical 
and systematic uncertainties. Bottom panel: The product of Bjorken energy density, $\epsilon_{Bj}$~\cite{Bjorken:1982qr}, and 
the formation time ($\tau$) in central heavy-ion collisions at mid-rapidity as a function of $\sqrt{s_{\rm {NN}}}$~\cite{Loizides:2011ys,Chatrchyan:2012mb,Mitchell:2012mx,Adler:2004zn,Adcox:2001ry,Aggarwal:2000bc}.}
  \end{figure}

Figure~\ref{fig:den} (top panel) shows the $\langle m_T \rangle$  values for pions in central heavy-ion collisions
as a function of $\sqrt{s_{\rm {NN}}}$. The $\langle m_T \rangle$ value increases with $\sqrt{s_{\rm {NN}}}$  at
lower AGS energies~\cite{Ahle:1999uy,Barrette:1999ry}, stays independent of $\sqrt{s_{\rm {NN}}}$ 
for the SPS energies~\cite{Afanasiev:2002mx,Alt:2007aa}  and then tends
to rise further with increasing $\sqrt{s_{\rm {NN}}}$ at the higher beam energies of  LHC. About 25\% 
increase in $\langle m_T \rangle$ is observed from RHIC~\cite{Abelev:2008ab,Abelev:2009bw}
 to LHC~\cite{Abelev:2012wca}. For a thermodynamic system, $\langle m_T \rangle$
can be an approximate representation of the temperature of the system, and dN/dy $\propto$ $ln(\sqrt{s_{\rm {NN}}})$ 
may represent its entropy~\cite{Mohanty:2003fy}. In such a scenario, the observations could reflect the characteristic signature of a 
phase transition, as proposed by Van Hove~\cite{vanhove}. Then, the constant value of  $\langle m_T \rangle$  vs. $\sqrt{s_{\rm {NN}}}$ 
has one possible interpretation in terms of formation of a mixed phase of a QGP and hadrons during the
evolution of the heavy-ion system. The energy domains accessed at RHIC and LHC will then correspond to 
partonic phase while those at AGS would reflect hadronic phase. However, there could be several other effects 
to which $\langle m_T \rangle$ is sensitive, which also need to be understood for proper interpretation of the data~\cite{Mohanty:2003fy}.

Figure~\ref{fig:den} (bottom panel) shows the product of the estimated Bjorken energy density 
($\epsilon_{Bj} = \frac{1}{A_{\perp}\tau} dE_{T}/dy$; $A_{\perp}$~\cite{Bjorken:1982qr} is the transverse overlap area of the nuclei and
$E_T$ is the transverse energy) and formation time ($\tau$) as a
function of $\sqrt{s_{\rm
    {NN}}}$~\cite{Loizides:2011ys,Chatrchyan:2012mb,Mitchell:2012mx,Adler:2004zn,Adcox:2001ry,Aggarwal:2000bc}. The
product of energy density and the formation time at LHC seems to be a
factor 3 larger compared to those attained at RHIC. 
If we assume the same value of $\tau_{0}$ (= 1
fm/$c$) for LHC and RHIC, the Bjorken energy density is about a factor
of 3 larger at the LHC
compared to that at RHIC in central collisions.

\subsubsection{Volume and decoupling time}

 %
  \begin{figure}
\includegraphics[scale=0.4]{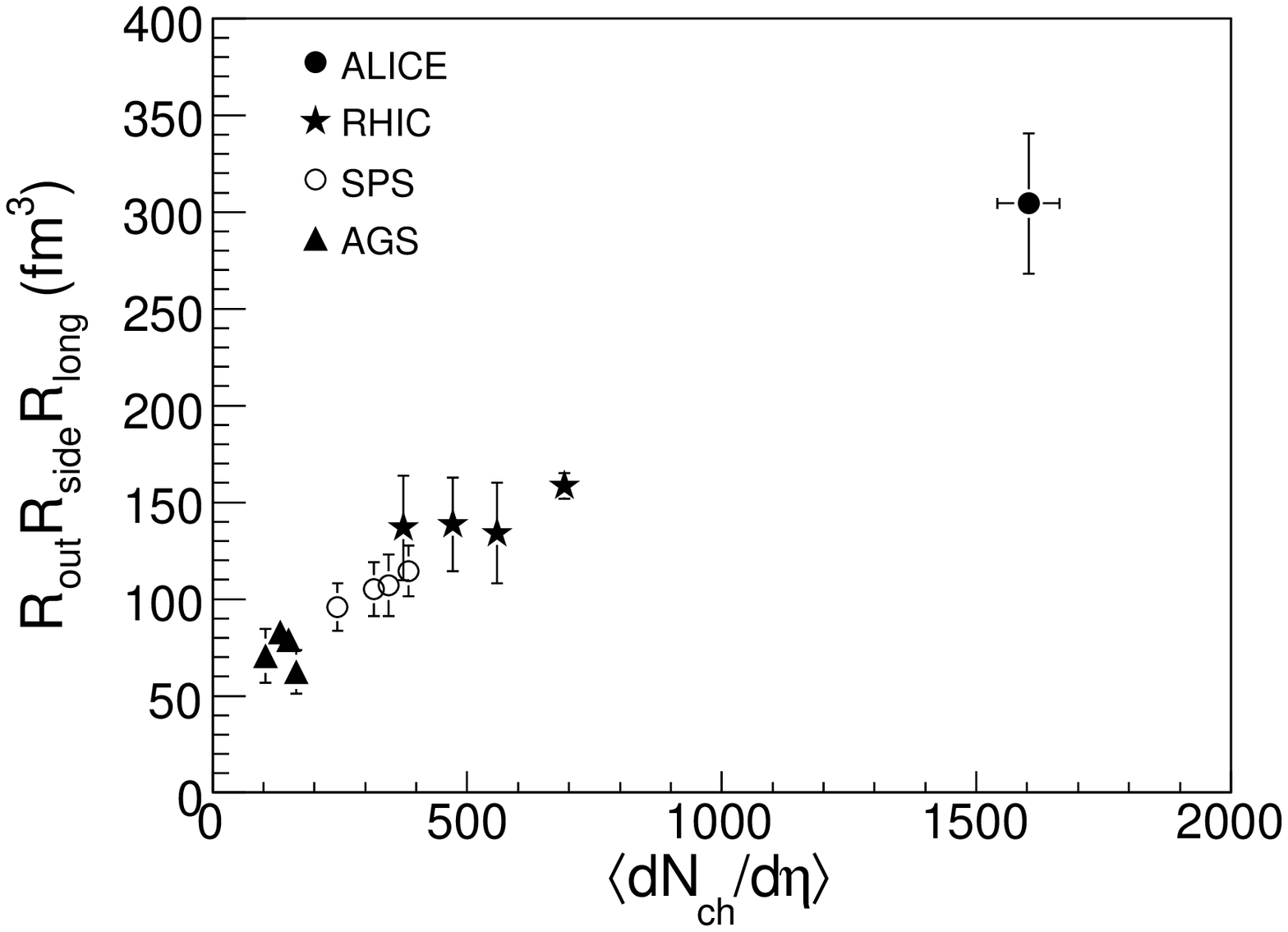}
\includegraphics[scale=0.4]{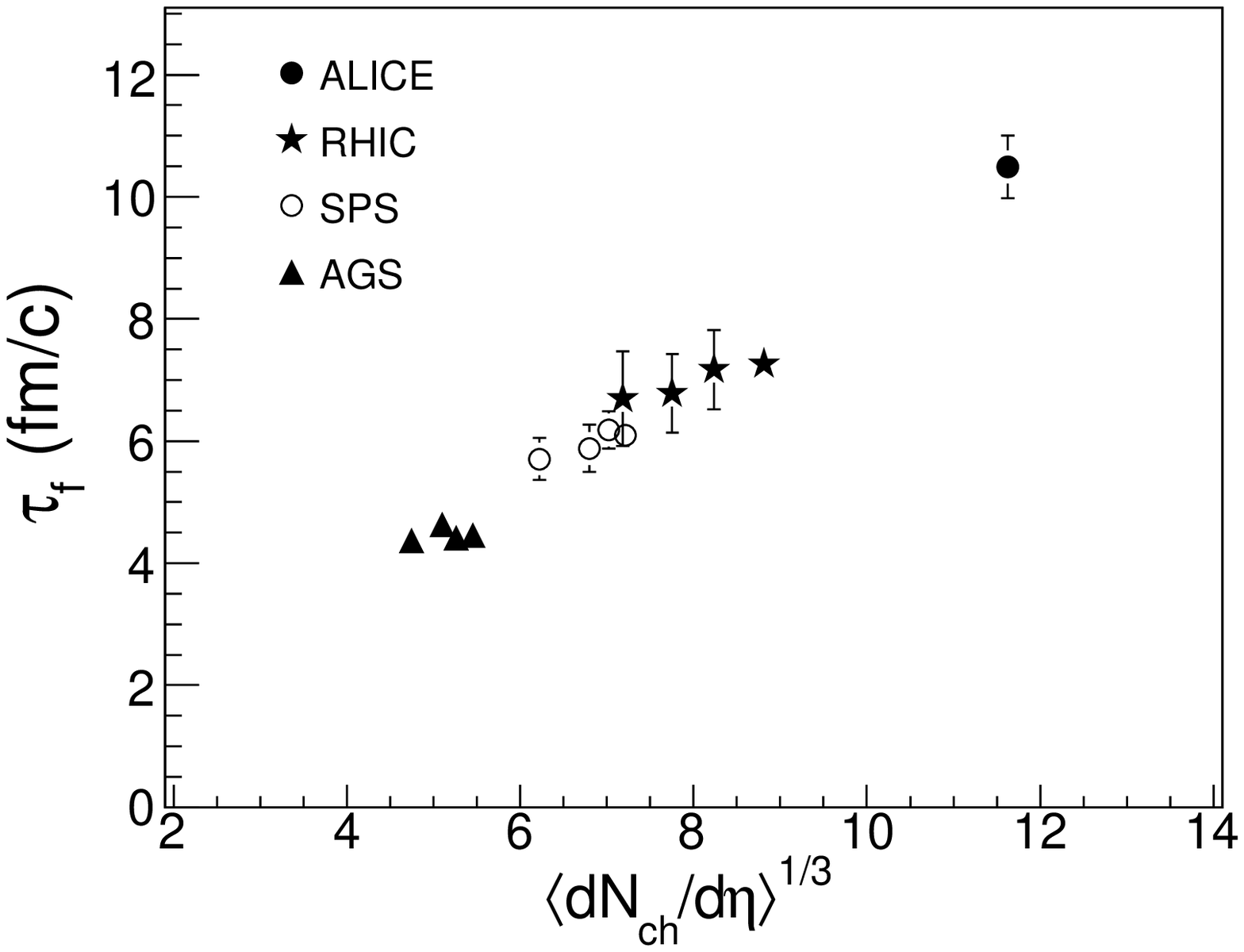}
\caption{\label{fig:hbt}
      Top panel: Product of the three pion HBT radii at $k_T$ (average transverse momenta of two pions) 
= 0.3 GeV/c for central heavy-ion collisions at AGS~\cite{Lisa:2000hw}, SPS~\cite{Alt:2007uj,Adamova:2002wi}, RHIC~\cite{Abelev:2009tp,Back:2004ug} 
and LHC~\cite{Aamodt:2011mr} energies. Bottom panel: The decoupling time extracted from 
$R_{\rm {long}}(k_{T})$ for central heavy-ion collisions at midrapidity at AGS, SPS, RHIC and LHC energies as a function of
 $(dN_{ch}/d\eta)^{1/3}$.}
  \end{figure}

The top panel of Fig.~\ref{fig:hbt} shows the energy dependence of the product of the three radii ($R_{\rm {out}}$, $R_{\rm {side}}$ and $R_{\rm {long}}$) obtained from pion
HBT, or Bose-Einstein correlation analysis. Here the ``out'' corresponds to the axis pointing along the pair transverse momentum, the ``side'' to the 
axis perpendicular to it in the transverse plane, and the ``long'' axis being 
along the beam (Bertsch-Pratt convention~\cite{Bertsch:1989vn,Pratt:1986cc}). 
The product of the radii is connected to the volume of the homogeneity region
at the last interaction. The product of the three radii shows a linear dependence on the charged-particle 
pseudorapidity density. The data indicates that the volume of  homogeneity region is two times larger at the LHC 
than at RHIC. 

Furthermore, within a hydrodynamic picture, the decoupling time for hadrons ($\tau_{f}$) at midrapidity can 
be estimated from the magnitude of radii $R_{\rm {long}}$ as: $R_{\rm {long}}^{2}$ =  
$\tau_{f}^{2} T K_{2}(m_{T}/T) / m_{T} K_{1}(m_{T}/T)$, with $m_{T} = \sqrt{m_{\pi}^{2} + k_{T}^{2}}$,
where $m_{\pi}$ is the mass of the pion, $T$ is the kinetic freeze-out
temperature 
and $K_{1}$ and $K_{2}$
are the integer order modified Bessel functions~\cite{Teaney:2003kp}. 
For the estimation of $\tau_{f}$, the average value of the kinetic freeze-out temperature $T$ is taken to be 120 MeV from AGS to LHC energies. 
However, the energy dependence of kinetic freeze-out temperature, as discussed in the next subsection, would provide a more
accurate description of the $\tau_{f}$ values.
The extracted $\tau_{f}$ values for central heavy-ion collisions at midrapidity 
at AGS~\cite{Lisa:2000hw}, SPS~\cite{Alt:2007uj,Adamova:2002wi}, RHIC~\cite{Abelev:2009tp,Back:2004ug} and
 LHC~\cite{Aamodt:2011mr} energies are shown as a function of cube-root of $dN_{ch}/d\eta$ in the bottom panel of Fig.~\ref{fig:hbt}. 
We observe that $\tau_{f}$ scales linearly with
$(dN_{ch}/d\eta)^{1/3}$ and is about 10 fm/$c$ at LHC energies. This 
value is about 40\% larger than at RHIC. 
It may be noted that the above expression ignores transverse
expansion of the system and finite chemical potential for pions. Also
there are uncertainties associated with freeze-out temperature that
could lead to variations in the extracted $\tau_f$ values.

\subsubsection{Freeze-out temperature and radial flow velocity}
 %
  \begin{figure}[htbp]
\includegraphics[scale=0.4]{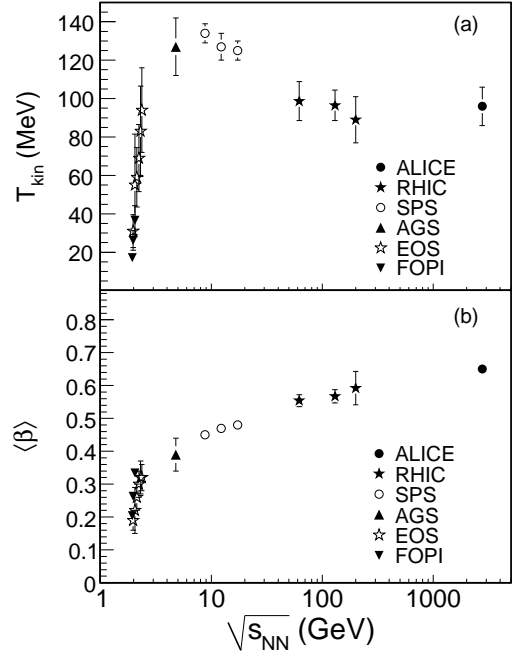}
    \caption{\label{fig:kfo}
      Kinetic freeze-out temperature (top panel) and radial flow
      velocity (bottom panel) in central heavy-ion collisions
as a function of collision energy.~\cite{Abelev:2008ab,Abelev:2012wca,Reisdorf:1996qj,Lisa:1994yr,Muntz:1997qt,Appelshauser:1997rr}}
  \end{figure}
The hadron yields and spectra reflects the properties of the bulk matter at chemical and kinetic freeze-out,
respectively. Generally, the point at which the inelastic collisions ceases is called the chemical freeze-out
and the point where even the elastic collisions stop is called the kinetic freeze-out. 

The transverse momentum
distribution of different particles contains two components, one random and the other collective. The random component
can be identified with the temperature of the system at kinetic freeze-out ($T_{\rm {kin}}$). The collective component,
which could arise from the matter density gradient from the center to the boundary of the fireball created in high
energy nuclear collisions is called collective flow in transverse direction ($\langle \beta \rangle$). Using the 
assumption that the system attains thermal equilibrium, the blast wave formulation can be used to extract $T_{\rm {kin}}$
and $\langle \beta \rangle$. These two quantities are shown in Fig.~\ref{fig:kfo} versus $\sqrt{s_{\rm {NN}}}$~\cite{Abelev:2008ab,Abelev:2012wca,Reisdorf:1996qj,Lisa:1994yr,Muntz:1997qt,Appelshauser:1997rr}. 
For beam energies at AGS and above, one observes a decrease in $T_{\rm {kin}}$ with $\sqrt{s_{\rm{NN}}}$. This indicates
that higher the beam energy, longer is the interactions among the constituents of the expanding system and lower the temperature.
From RHIC top energy to LHC there seems to be however a saturation in the value of $T_{\rm {kin}}$. In contrast to 
the temperature, the collective flow increases with increase in beam energy rapidly, reaching a value close to 0.6
times the speed of light at the LHC energy. 
 %
  \begin{figure}[htbp]
\includegraphics[scale=0.4]{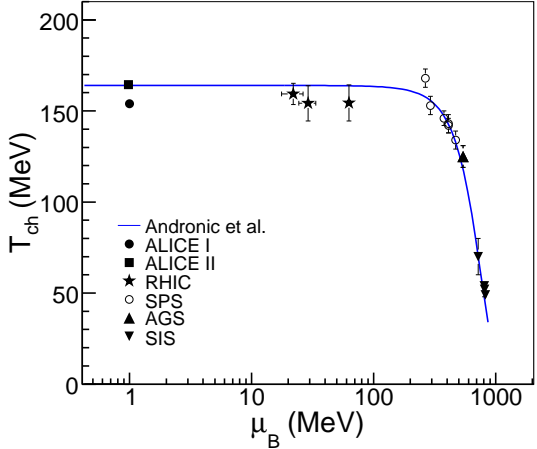}
    \caption{\label{fig:chfo}
      (Color online) Chemical freeze-out temperature versus baryon chemical potential in central heavy-ion collisions~\cite{Abelev:2008ab,Abelev:2012wca,Andronic:2005yp,Andronic:2008gu,Cleymans:1998yf,Cleymans:1997sw,BraunMunzinger:1994xr,BraunMunzinger:1999qy,Becattini:2002ix,Becattini:2000jw}.
The curve corresponds to model calculations from Refs.~\cite{Andronic:2005yp,Andronic:2008gu}.}
  \end{figure}

Figure~\ref{fig:chfo} shows the chemical freeze-out temperature ($T_{\rm {ch}}$) versus the baryon chemical potential
($\mu_{\rm B}$) in central heavy-ion collisions~\cite{Abelev:2008ab,Abelev:2012wca,Andronic:2005yp,Andronic:2008gu,Cleymans:1998yf,Cleymans:1997sw,BraunMunzinger:1994xr,BraunMunzinger:1999qy,Becattini:2002ix,Becattini:2000jw}. These quantities are obtained by fitting the particle yields to a
statistical model assuming thermal equilibrium within the framework of
a Grand Canonical ensemble. 
There are two values of 
temperature quoted for LHC energies. A $T_{\rm ch}$ value of about 164 MeV and fixed $\mu_{\rm B}$ value of 1 MeV seems to 
reproduce the multi-strange ratios (involving $\Xi$ and $\Omega$) quite well, but were observed to
miss the data for $p/\pi$ and $\Lambda/\pi$. On the other hand, the statistical thermal model prediction with $T_{ch}$ = 152 MeV 
and fixed  $\mu_{\rm B}$ = 1 MeV fits the measured $p/\pi$ and $\Lambda/\pi$ ratios better but misses the 
ratios involving multi-strange hadrons~\cite{Milano:2013sza}. This issue is not yet resolved, with being possibly related to 
hadronic final state interactions~\cite{Steinheimer:2012rd}. 
The curve corresponds to generalization of the
energy dependence of  $T_{\rm{ch}}-\mu_B$ using statistical thermal model
  calculations~\cite{Andronic:2005yp,Andronic:2008gu}. The model works within
the framework of a Grand Canonical ensemble and 
  takes as input the produced particle yields from experiments to extract
  the freeze-out parameters such as $T_{\rm{ch}}$ and
$\mu_B$.

\subsubsection{Fluctuations}

 %
  \begin{figure}[htbp]
\includegraphics[scale=0.4]{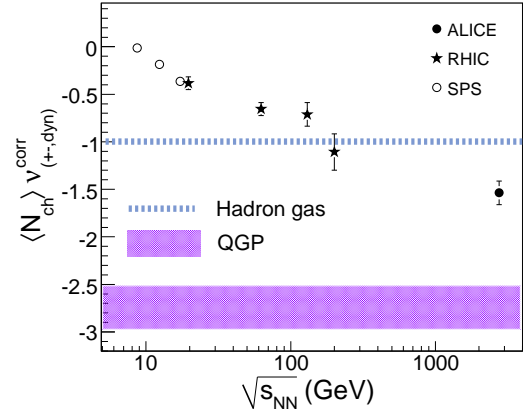}
    \caption{\label{fig:fluct}
      (Color online) Energy dependence of net-charge fluctuations about midrapidity in central heavy-ion collisions at 
SPS~\cite{Sako:2004pw}, RHIC~\cite{Abelev:2008jg} and LHC~\cite{Abelev:2012pv} energies. Also shown are the expectations from a hadron resonance gas model and for a simple QGP picture~\cite{Jeon:2000wg}.}
  \end{figure}
One of the proposed signatures to search for the phase transition from hadronic to partonic medium,
is to study the net-charge fluctuations in heavy ion collisions. The
partonic phase has constituents
with fractional charges while the hadronic phase has constituents with integral units of charge, hence the measure
of the fluctuations in the net-charge particle production is expected to be different in these two cases.
Specifically, net-charge fluctuations are expected to be smaller if the system underwent a phase transition.
However it is important to address that how these fluctuations may or may not survive the evolution
of the system in the heavy-ion collisions. An experimental
measure of net-charge fluctuations is defined as $\nu(+-,dyn)$ = $\frac {\langle N_{+}(N_{+}-1)\rangle}{\langle N_{+}^{2}\rangle}$
+ $\frac {\langle N_{-}(N_{-}-1)\rangle}{\langle N_{-}^{2}\rangle}$ - 2  $\frac{\langle N_{-}N_{+}\rangle}{\langle N_{-}\rangle \langle N_{+}\rangle}$, where $\langle N_{-}\rangle$ and $\langle N_{+}\rangle$ are average negative and positive 
charged particle multiplicity, respectively~\cite{Pruneau:2002yf}.

Figure~\ref{fig:fluct} shows the product of $\nu(+-,dyn)$ and $\langle N_{ch}\rangle$ (average number of charged particles) as 
a function of $\sqrt{s_{\rm {NN}}}$~\cite{Abelev:2012pv,Abelev:2008jg,Sako:2004pw}.  
We find that this fluctuation observable rapidly decreases with $\sqrt{s_{\rm {NN}}}$
and approaches expectation for a simple QGP-like scenario~\cite{Jeon:2000wg} as we move from RHIC to LHC energies. 
Given that several other observables already indicate that a hot and dense medium of color charges has been formed at RHIC and LHC,
the net-charge fluctuation result may indicate that the observable  $\nu(+-,dyn)$ is not sensitive enough to QGP physics or the process of hadronization
washes out the QGP signal for this observable. 
It may be also noted that the models results do not incorporate
the acceptance effects and do not consider any dynamic evolution of the system like for example the dilution of the
signals in the hadronization process.

\subsection{Azimuthal anisotropy}

\subsubsection{Energy dependence of $p_{T}$ integrated $v_{2}$}

  \begin{figure}[htbp]
\includegraphics[scale=0.4]{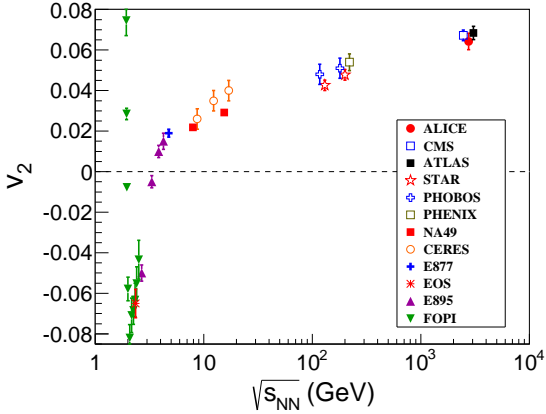}
    \caption{\label{fig:v2edep}
      (Color online) Transverse momentum integrated $v_{2}$ close to midrapidity for charged (Z = 1) particles
      for collision centralities around 20-30\% as a function of center of mass energy. }
  \end{figure}

Figure~\ref{fig:v2edep} shows the $p_{T}$ integrated $v_{2}$ close to midrapidity of charged particles
for collision centralities around 20-30\% as a function of center of mass energy. We observe
that there is an increase in magnitude of $v_{2}$ by about 30\% from top RHIC energy ($\sqrt{s_{\rm {NN}}}$ = 200 GeV) to
LHC energy ($\sqrt{s_{\rm {NN}}}$ = 2.76 TeV). This needs to be viewed within the context of a similar magnitude
of increase in $\langle p_{T} \rangle$ of pions from RHIC to LHC energies. The increase of $v_{2}$  beyond beam energy of 
10 AGeV is logarithmic in $\sqrt{s_{\rm {NN}}}$. This is expected to be determined by the pressure 
gradient-driven expansion of the almond-shape fireball produced in the initial stages of a non-central
heavy-ion collision~\cite{Ollitrault:1992bk}. While for 
 $v_{2}$ measured at lower beam energies, the dependences observed is due to interplay of 
passing time of spectators and time scale of expansion of the system. A preference for an in-plane  
emission versus out-of-plane ( ``squeeze-out'') pattern of particles as a function of beam energy
is observed. The experimental data used are from FOPI~\cite{Andronic:2004cp,Bastid:2005ct}, EOS, E895~\cite{BraunMunzinger:1998cg}, E877~\cite{Pinkenburg:1999ya}, CERES~\cite{Adamova:2002qx}, NA49~\cite{Alt:2003ab}, STAR~\cite{Adams:2004bi}, PHOBOS~\cite{Back:2004zg}, PHENIX~\cite{Afanasiev:2009wq}, 
ALICE~\cite{Aamodt:2010pa}, ATLAS~\cite{ATLAS:2012gna} and CMS~\cite{Chatrchyan:2012xq} 
experiments. Charged particles are used for LHC, RHIC, CERES and E877 
experiments, pion data is used from
NA49 experiment, protons results are from EOS and E895 experiment and FOPI results are for all particles with Z=1.

\subsubsection{Azimuthal anisotropy coefficients versus transverse momentum}

 %
  \begin{figure}[htbp]
\includegraphics[scale=0.45]{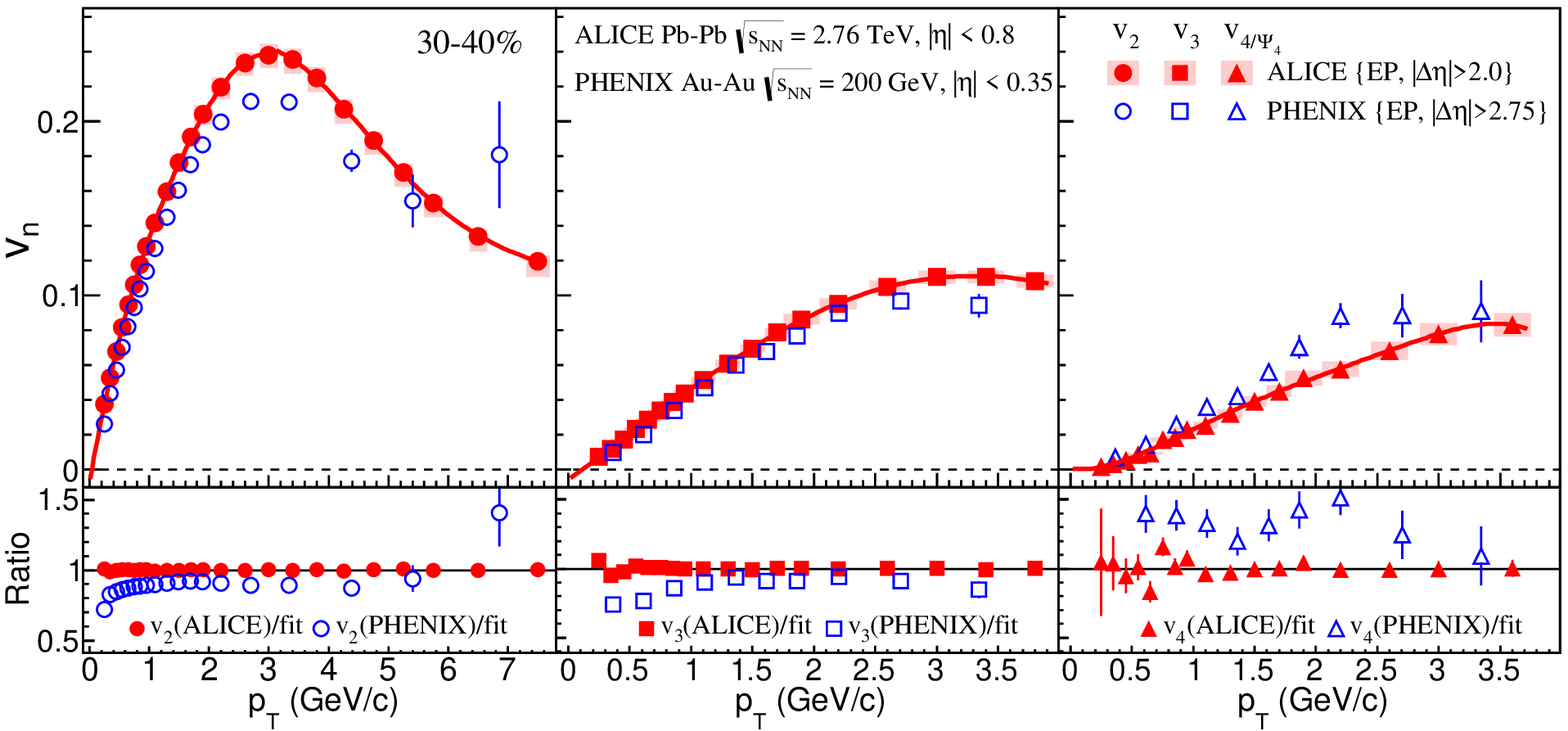}
\includegraphics[scale=0.45]{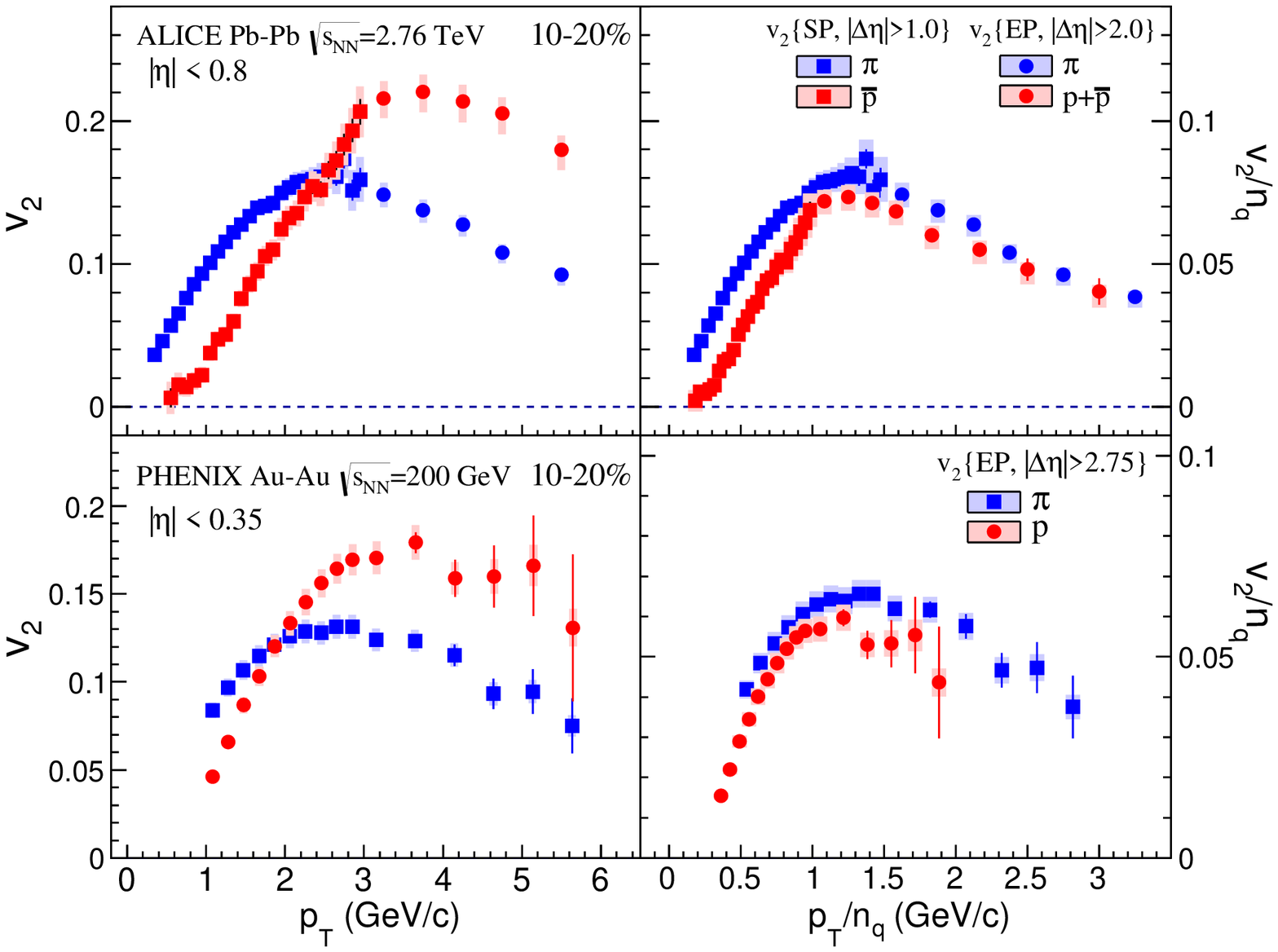}
    \caption{\label{fig:v2rhiclhc}
      (Color online) Top:  Comparison of $v_{n}(p_{T})$ at midrapidity for 30-40\% collision centrality at RHIC
(Au+Au collisions at $\sqrt{s_{\rm {NN}}}$ = 200 GeV from PHENIX experiment~\cite{Adare:2011tg}) and at LHC (Pb+Pb collisions 
at $\sqrt{s_{\rm {NN}}}$~=~2.76 TeV from ALICE experiment~\cite{ALICE:2011ab}). The bottom panels show the ratio of $v_{n}$ at LHC and RHIC.
Bottom: $v_{2}$ versus $p_{T}$ and $v_{2}/n_{q}$ versus $p_{T}/n_{q}$ for pions and protons at midrapidity for 10-20\%
collision centrality from Au+Au collisions at $\sqrt{s_{\rm {NN}}}$ = 200 GeV (PHENIX experiment~\cite{Adare:2012vq}) and Pb+Pb collisions
at $\sqrt{s_{\rm {NN}}}$ = 2.76 TeV (ALICE experiment~\cite{Abelev:2012di}). }
  \end{figure}

Figure~\ref{fig:v2rhiclhc} (top) shows the comparison of $v_{2}(p_{T})$, $v_{3}(p_{T})$ and $v_{4}(p_{T})$ 
for 30-40\% collision centrality at RHIC (PHENIX experiment~\cite{Adare:2011tg}) and LHC (ALICE~\cite{ALICE:2011ab}) at midrapidity in Au+Au and Pb+Pb collisions, respectively.
The bottom panel of this figure shows the ratio of LHC and RHIC results to a polynomial fit to the LHC data. 
The $v_{n}(p_{T})$ measurement techniques are similar at RHIC and LHC energies. One observes that at lower $p_{T}$
( $<$ 2 GeV/$c$) the  $v_{2}(p_{T})$ and $v_{3}(p_{T})$ are about 10-20\% smaller at RHIC compared to the corresponding LHC
results. However at higher $p_{T}$ the results are quite similar. The $v_{4}(p_{T})$ seems higher at RHIC compared to LHC.

One of the most striking observations to come out from RHIC is the number of constituent quark ($n_{q}$) scaling of
$v_{2}(p_{T})$ for identified hadrons. The basis of such a scaling is the splitting of $v_{2}(p_{T})$ between baryons
and mesons at intermediate  $p_{T}$ (2--6~GeV/$c$). This is shown in
the bottom panels of the Fig.~\ref{fig:v2rhiclhc} (bottom).
Such a splitting between baryon and meson $v_{2}(p_{T})$ is also observed at intermediate $p_{T}$  at LHC energies 
(seen in the top panels of the bottom Fig.~\ref{fig:v2rhiclhc}). However the degree to which $n_{q}$ scaling holds 
could be different at RHIC~\cite{Adare:2012vq} and LHC~\cite{Abelev:2012di} energies. The $n_{q}$ scaling is much more
closely followed at RHIC compared to LHC. It may be noted that there are several factors which could dilute such scalings,
which includes energy dependence of radial flow, an admixture of higher Fock states and consideration of 
a realistic momentum distribution of quarks inside a hadron~\cite{Muller:2005pv,Greco:2005jk}. 
The observation of the baryon-meson splitting is commonly interpreted as
due to substantial amount of collectivity being generated in the de-confined phase. Another important feature is that at both 
RHIC and LHC energies a clear hydrodynamic feature of mass dependence of $v_{2}(p_{T})$ is observed at low  $p_{T}$ ($<$ 2 GeV/$c$).

 %
  \begin{figure}
\includegraphics[scale=0.4]{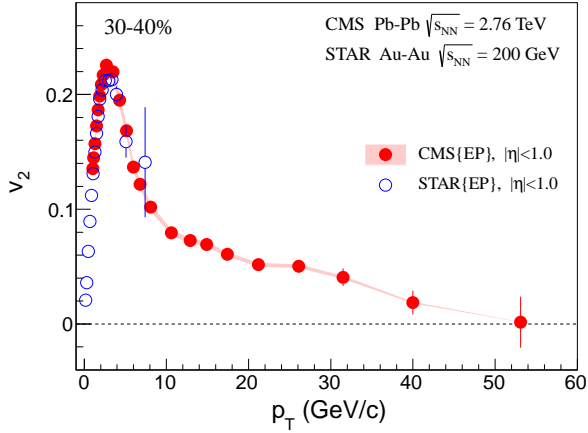}
    \caption{\label{fig:v2highpt}
      (Color online) Comparison of $v_{2}(p_{T})$ at midrapidity for 30-40\% collision centrality at RHIC (Au+Au collisions
at $\sqrt{s_{\rm {NN}}}$ = 200 GeV from STAR experiment) and at LHC (Pb+Pb collisions at $\sqrt{s_{\rm {NN}}}$ = 2.76 TeV from
CMS experiment~\cite{Chatrchyan:2012xq}). The shaded band about CMS data point are systematic errors and vertical lines represent statistical errors. }
  \end{figure}

Figure~\ref{fig:v2highpt} shows the charged hadron $v_{2}(p_{T})$ for 30-40\% collision centrality in Au+Au collisions at 
 \sqrts{200} and Pb+Pb collisions at  $\sqrt{s_{\rm {NN}}}$ = 2.76 TeV for $|\eta|$ $<$ 1~\cite{Chatrchyan:2012xq}. This figure
demonstrates the kinematic reach for higher energy collisions at LHC relative to RHIC. LHC data allows us to study
the  $v_{2}(p_{T})$ in the $p_{T}$ range never measured before in heavy-ion collisions. 
The $v_{2}(p_{T})$ $\sim$ 0
for $p_{T}$~$>$~40~GeV/$c$ might suggests that those particles must have been emitted very early in the interactions
when the collective effects had not set in. These high transverse momentum data is useful to understand 
the effects of the initial geometry or path-length dependence of various properties associated with parton modification inside the hot QCD medium.
In addition, it also provides significantly improved precision measurement of $v_{2}$ for 12 $<$ $p_{T}$ $<$ 20 GeV/$c$.

\subsubsection{Flow fluctuations}

 %
  \begin{figure}[htbp]
\includegraphics[scale=0.4]{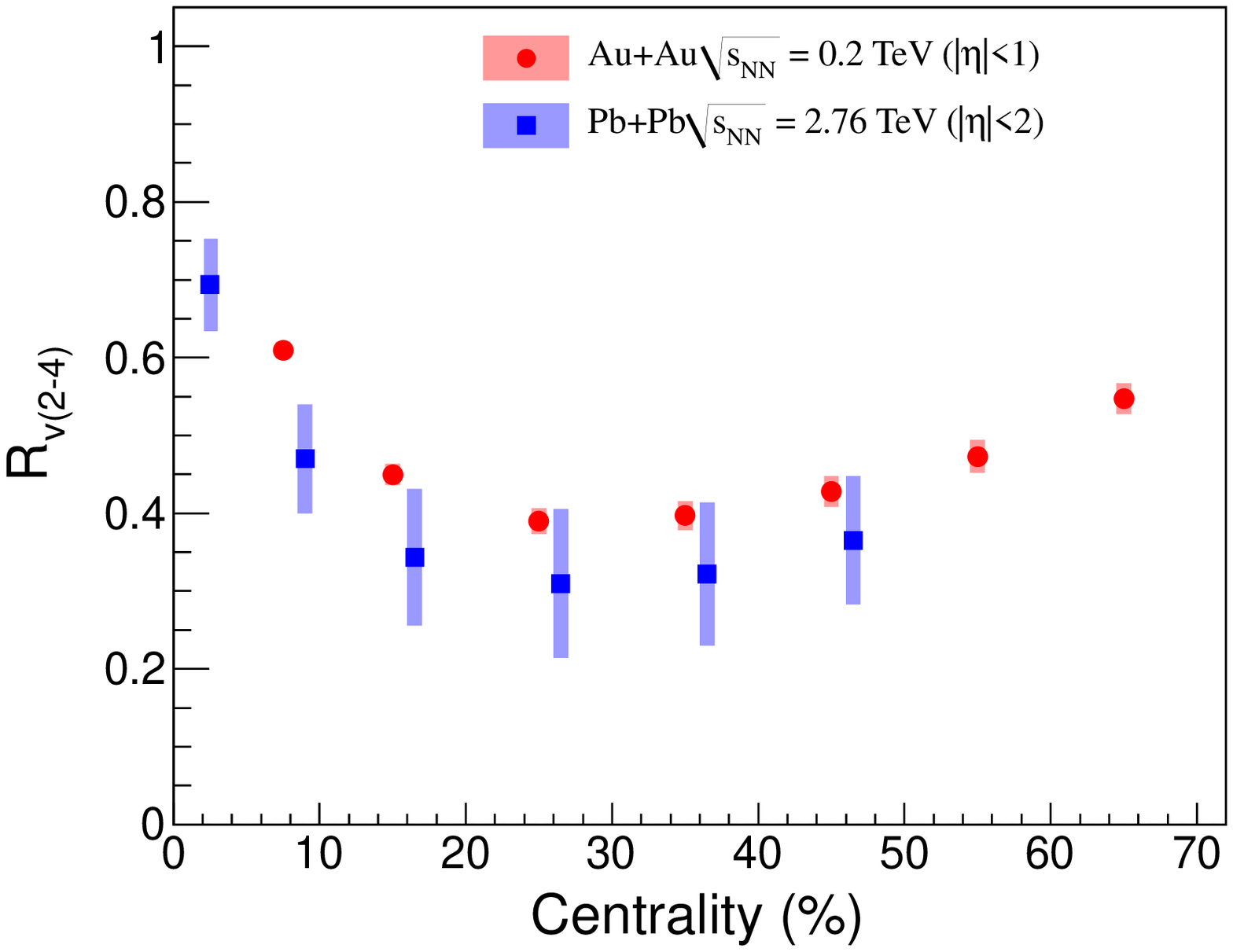}
\includegraphics[scale=0.4]{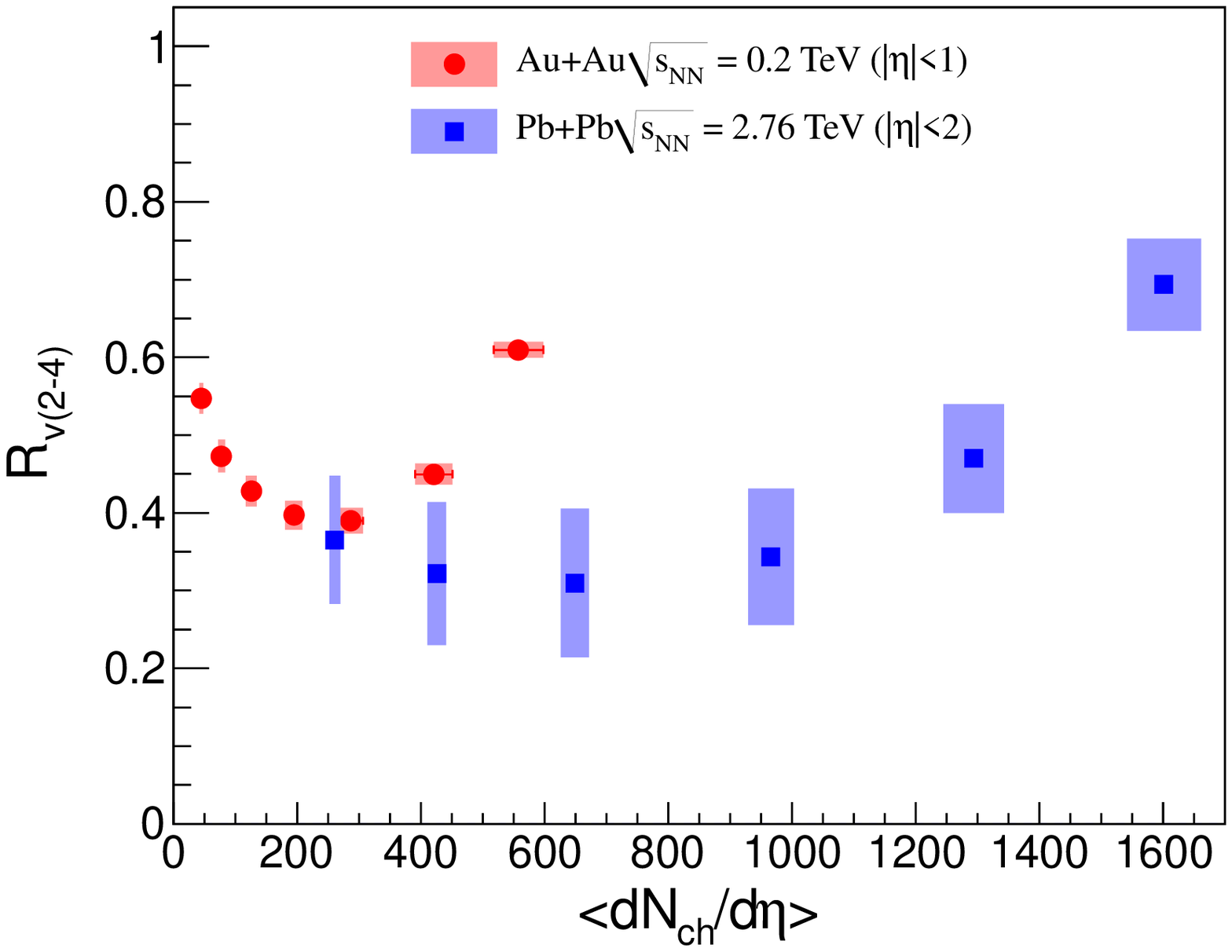}
    \caption{\label{fig:v2fluc}
      (Color online) 
The ratio $R_{v(2-4)}=\sqrt{(v_{2}\{2\}^{2} - v_{2}\{4\}^{2})/(v_{2}\{2\}^{2} + v_{2}\{4\}^{2})}$, an estimate of
$v_{2}$ fluctuations plotted as a function of collision centrality
(top panel) and $\langle dN_{ch}/d\eta \rangle$ (bottom panel)
for RHIC (STAR experiment: Au+Au collisions at $\sqrt{s_{\rm {NN}}}$ = 200 GeV~\cite{Agakishiev:2011eq}) 
and LHC (ALICE: Pb+Pb collisions at $\sqrt{s_{\rm {NN}}}$ = 2.76 TeV~\cite{Abelev:2012di}) at midrapidity. The bands reflect the systematic errors. }
  \end{figure}
Fluctuations in azimuthal anisotropy coefficient $v_{2}$ have gained quite an attention in recent times. In particular, 
the measurement of event-by-event  $v_{2}$ fluctuations can pose new constraints on the models of the initial state of the 
collision and its subsequent hydrodynamic evolution. In extracting  event-by-event  $v_{2}$ fluctuations one needs to
separate non-flow effects and so far there is no direct method to decouple $v_{2}$ fluctuations  and non-flow effects
in a model independent way from the experimental measurements. However, several techniques exists where the non-flow
effects can be minimized, for example, flow and non-flow contributions can be possibly separated to a great extent 
with a detailed study of two particle correlation function in $\Delta \phi$ and its dependence on $\eta$ and $\Delta \eta$.
Here we discuss another technique to extract and compare the $v_{2}$ fluctuations at RHIC and LHC. We assume that the difference
between $v_{2}\{2\}$ (two particle cumulant)  and $v_{2}\{4\}$ (four particle cumulant) is dominated by  $v_{2}$ fluctuations
and non-flow effect is negligible for $v_{2}\{4\}$. Then the ratio $R_{v(2-4)}$ = $\sqrt{(v_{2}\{2\}^{2} - v_{2}\{4\}^{2})/(v_{2}\{2\}^{2} + v_{2}\{4\}^{2})}$ can be considered as an estimate for $v_{2}$ fluctuations in the data. 
Figure~\ref{fig:v2fluc} shows the $R_{v(2-4)}$ as a function of
collision centrality and $\langle dN_{ch}/d\eta \rangle$ for RHIC~\cite{Agakishiev:2011eq} 
and LHC~\cite{Abelev:2012di} energies. 
The centrality dependence of $R_{v(2-4)}$ at RHIC or LHC as seen in Fig.~\ref{fig:v2fluc} could be an interplay of residual
non-flow effects which increases for central collisions and multiplicity fluctuations which dominates smaller systems.
It is striking to see that $R_{v(2-4)}$ when presented as a function of \% cross section is similar at RHIC and
LHC suggesting it reflects features associated with initial state of the collisions for example the event-by-event fluctuations in 
the eccentricity of the system. But when presented as a function of $dN_{ch}/d\eta$ it tends to suggest a different behaviour
for most central collisions at RHIC.

\begin{widetext}

 %
  \begin{figure}[htbp]
\includegraphics[scale=0.48]{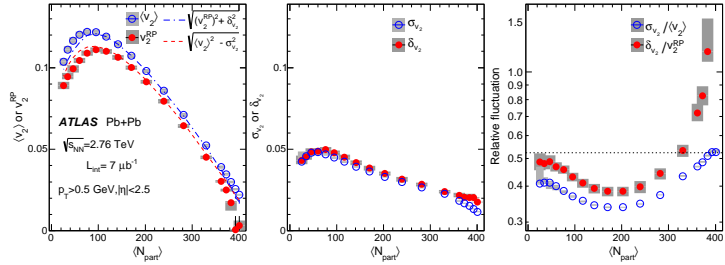}
\caption{\label{fig:fit2}
      (Color online) 
The dependence of $v_2^{\mathrm{RP}}$ and $\langle v_2\rangle$ (left
panel),  $\delta_{_{v_2}}$ and $\sigma_{v_2}$ (middle panel), and $\delta_{_{v_2}}/v_2^{\mathrm{RP}}$ and $\sigma_{_{v_2}}/\langle v_2\rangle$ (right panel) on $\langle N_{\mathrm{part}}\rangle$~\cite{Aad:2013xma}. The shaded boxes indicate the systematic uncertainties.
}
  \end{figure}

Recently, a great interest has been generated on extracting initial condition and flow fluctuation information from the measurement 
of the probability distribution of $v_{n}$ at LHC. The probability density of ${v}_n$ can be expressed as a Gaussian function 
in transverse plane~\cite{Voloshin:2007pc} as 
$p({v}_n)~=~\frac{1}{2\pi\delta^2_{_{v_n}}} e^{-\left({v}_n-{v}_n^{\mathrm{\;RP}}\right)^2 \big{/}\left(2\delta^2_{_{v_n}}\right)}$ 
or as one dimensional Bessel-Gaussian
function~\cite{Voloshin:1994mz,Voloshin:2008dg} as 
$p(v_n) =
\frac{v_n}{\delta_{_{v_n}}^2}e^{-\frac{(v_n)^2+(v_n^{\mathrm{RP}})^2}{2\delta_{_{v_n}}^2}}
I_0\left(\frac{v_n^{\mathrm{RP}}v_n}{\delta_{_{v_n}}^2}\right)$. 
Where  $I_0$ is the modified Bessel function of the first kind, $\delta_{v_n}$ is the 
fluctuation in $v_{n}$. With $ \delta_{_{v_n}} \approx \sigma_{v_n}$
for $\delta_{v_n} \ll v_n^{\mathrm{RP}}$ ($v_{n}$ measured with respect to reaction plane).

Figure~\ref{fig:fit2} shows the $v_2^{\mathrm{RP}}$ and $\delta_{_{v_2}}$ values extracted from the $v_2$ distributions as a 
function of $\langle N_{\mathrm{part}}\rangle$ by fitting to the above probability functions~\cite{Aad:2013xma}. They are compared with 
values of $\langle v_2\rangle$ and $\sigma_{v_2}$ obtained 
directly from the $v_2$ distributions. The $v_2^{\mathrm{RP}}$ value is always smaller than the value for $\langle v_2\rangle$, 
and it decreases to zero in the 0--2\% centrality interval. The value of $\delta_{_{v_2}}$ is close to $\sigma_{_{v_2}}$ except 
in the most central collisions. This leads to a value of $\delta_{_{v_2}}/v_2^{\mathrm{RP}}$ larger than 
$\sigma_{_{v_2}}/\langle v_2\rangle$ over the full centrality range as shown in the right panel of figure~\ref{fig:fit2}. The value 
of $\delta_{_{v_2}}/v_2^{\mathrm{RP}}$ decreases with $\langle N_{\mathrm{part}}\rangle$ and reaches a minimum at $\langle N_{\mathrm{part}}\rangle\approx200$,
but then increases for more central collisions. Thus, the
event-by-event $v_{2}$ distribution brings additional insight for the understanding
of $v_{2}$ fluctuations.

\end{widetext}

\subsection{Nuclear modification factor}

\begin{widetext}

 %
  \begin{figure}
\includegraphics[scale=0.8]{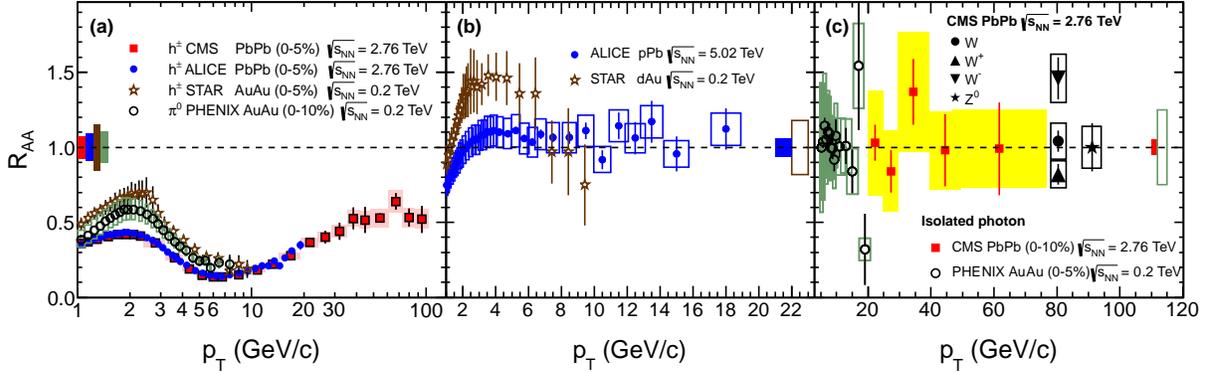}
    \caption{\label{fig:fig13}
     (Color online) (a) Nuclear modification factor $R_{\rm {AA}}$ of charged hadrons measured by ALICE~\cite{Aamodt:2010jd} and CMS~\cite{CMS:2012aa} 
experiments at midrapidity for 0-5\% most central Pb+Pb collisions at $\sqrt{s_{\mathrm {NN}}}$ = 2.76 TeV. 
For comparison, shown are the $R_{\rm {AA}}$ of charged hadrons at midrapidity for 0-5\% most central collisions
measured by STAR~\cite{Adams:2003im} and $R_{\rm {AA}}$ of $\pi^{0}$ at midrapidity for 0-10\% most central collisions
measured by PHENIX~\cite{Adler:2006bw} for Au+Au collisions at $\sqrt{s_{\mathrm {NN}}}$ = 200 GeV.
(b) Comparison of nuclear modification factor for charged hadrons versus $p_{T}$ 
at midrapidity for minimum bias collisions in $d$+Au collisions at  
$\sqrt{s_{\mathrm {NN}}}$ = 200 GeV~\cite{Adams:2003im} and $p$+Pb collisions at  $\sqrt{s_{\mathrm {NN}}}$ = 5.02 TeV~\cite{ALICE:2012mj}. 
(c) The nuclear modification factor  versus $p_{T}$ for isolated photons in central 
nucleus-nucleus collisions at $\sqrt{s_{\mathrm {NN}}}$ = 200 GeV~\cite{Adler:2005ig} and 2.76 TeV~\cite{Chatrchyan:2012vq}. Also shown
are the $p_{T}$ integrated $R_{\rm {AA}}$ of $W^{\pm}$~\cite{Chatrchyan:2012nt} and $Z$ bosons~\cite{Chatrchyan:2011ua} at 
corresponding $m_{T}$ at LHC energies.  Open and shaded boxes represents the systematic uncertainties in the experimental
measurements and normalization uncertainties respectively.}
  \end{figure}

Figure~\ref{fig:fig13} shows the  $R_{\rm {AA}}$ of various particles produced in heavy-ion collisions at RHIC and LHC.
In Fig.~\ref{fig:fig13}(a), we observe that the shape of the $R_{\rm {AA}}$ versus $p_{T}$ of charged hadrons at RHIC 
and LHC~\cite{Aamodt:2010jd,CMS:2012aa} are very similar for the common $p_{T}$ range of measurements. The values $R_{\rm {AA}}$ at RHIC are higher
compared to those at LHC energies up to  $p_{T}$ $<$ 8 GeV/$c$. The higher kinematic reach of LHC in $p_{T}$ allows
us to see the full $p_{T}$ evolution of $R_{\rm {AA}}$ in high energy
heavy-ion collisions. All these measurements suggest that the 
energy loss of partons in the medium formed in heavy-ion collisions at LHC energies is perhaps larger compared to those
at RHIC. In Fig.~\ref{fig:fig13}(b), we observe that the nuclear modification factors for $d$+Au collisions at
$\sqrt{s_{\mathrm {NN}}}$~=~200 GeV~\cite{Adams:2003im} and  $p$+Pb collisions at  $\sqrt{s_{\mathrm {NN}}}$ = 5.02 TeV~\cite{ALICE:2012mj} are greater 
than unity for the $p_{T}$ $>$ 2 GeV/$c$. The values for RHIC are slightly larger compared to those for LHC.
A value greater than unity for the nuclear modification factor in $p(d)$+A collisions are generally interpreted 
as due to Cronin effect~\cite{Antreasyan:1977aw,Antreasyan:1977au}. However, several other physics effects could influence the magnitude of the 
nuclear modification factor in  $p(d)$+A collisions such as nuclear shadowing and gluon saturation effects.
But the results that the nuclear modification factors in $p(d)$+A
collisions are not below unity, strengthen
the argument (from experimental point of view) that a hot and dense
medium of color charges is formed in A+A collisions at RHIC and LHC. 
In Fig.~\ref{fig:fig13}(c), we show the $R_{\rm {AA}}$ of particles
than do not participate in strong interactions and 
some of them are most likely formed in the very early stages of the collisions. These particles (photon~\cite{Adler:2005ig,Chatrchyan:2012vq}, $W^{\pm}$~\cite{Chatrchyan:2012nt} and $Z$~\cite{Chatrchyan:2011ua}  bosons)
have a $R_{\rm {AA}}$ $\sim$ 1, indicating that the $R_{\rm {AA}}$ $<$
1, observed for charged hadrons in A+A collisions, are due to the strong 
interactions in a dense medium consisting of color charges. 

\end{widetext}

\section{Comparison to model calculations}
\label{sec:model}

In this section, we compare some of the experimental observables discussed above with corresponding model
calculations. This helps us to interpret the data at both RHIC and LHC
energies. 
We restrict our discussion on the comparison of the models with the experimental data for charged particle
production, ratio of kaon to pion yields as a function of beam energy, $p_{\rm T}$ dependence of $v_2$ and $R_{\rm {AA}}$
for charged particles and pions. For the charged particle production, we compare the experimental data with models 
inspired by the perturbative QCD-based calculations (HIJING, DPMJET), with macroscopic models (statistical and hydrodynamical), 
with microscopic models (string, transport, cascade, etc) and calculations which are derived by the different parametrizations 
of the nucleon-nucleon and nucleus-nucleus lower energy data. The ratio of kaon to pion yields for different beam energies 
are compared with the statistical and thermal models. The transverse momentum dependence of $v_2$ are compared with models 
incorporating the calculations based on hydrodynamic and transport approaches. Finally, the $R_{\rm AA}$ results are compared
with the perturbative QCD based calculations with different mechanism for the parton energy loss in the presence of colored medium.

\subsection{Charged particle multiplicity density and particle ratio}

  \begin{figure}[htbp]
\includegraphics[scale=0.4]{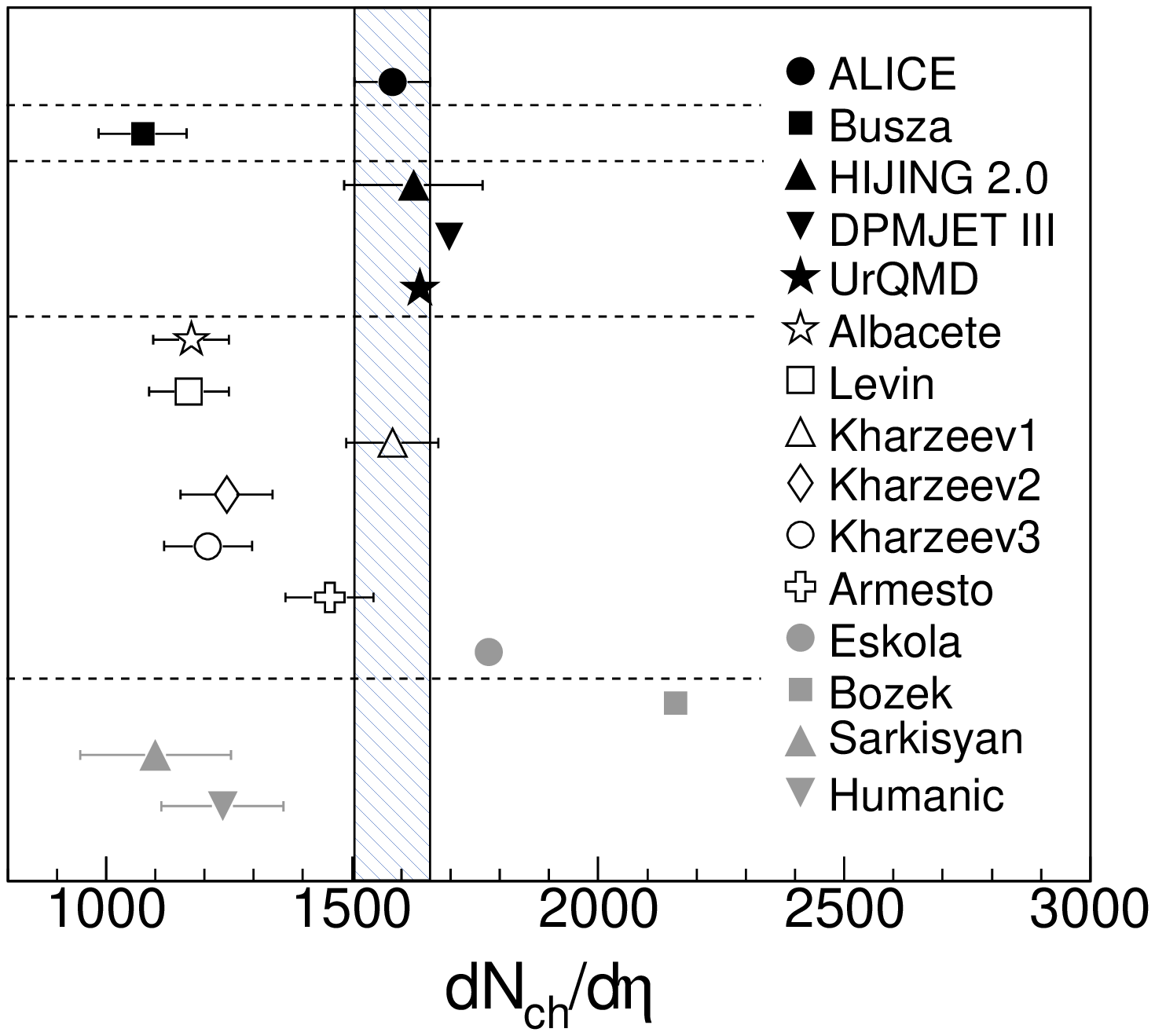}
\includegraphics[scale=0.4]{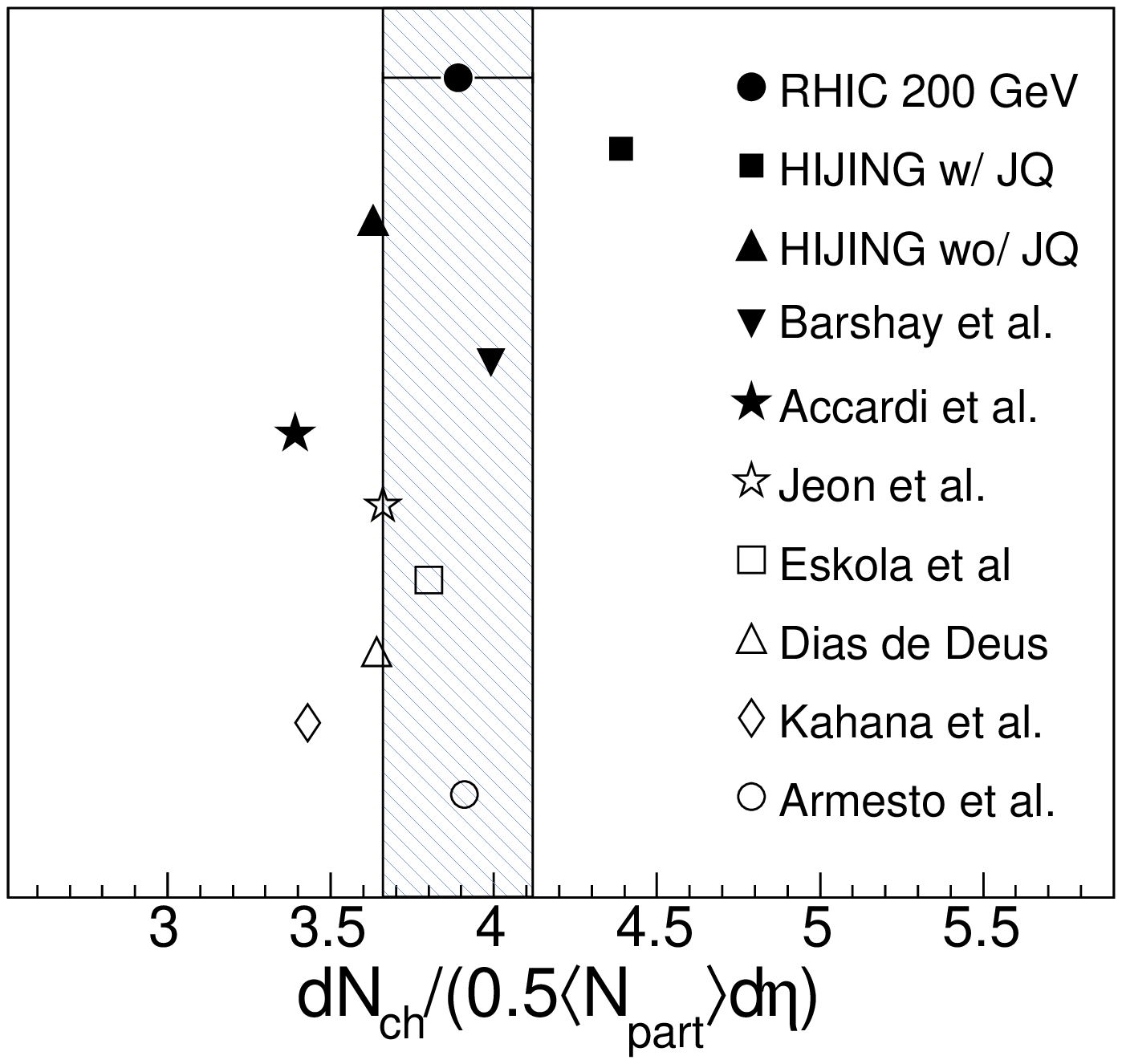}
    \caption{\label{fig:fig14}
      (Color online) Comparison of $dN_{\rm {ch}}/d\eta$ measurement at midrapidity for central heavy-ion collisions
at RHIC and LHC with model predictions.}
  \end{figure}

Figure~\ref{fig:fig14} compares the measured charged particle pseudorapidity density
at RHIC (0.2 TeV) and LHC (2.76 TeV) energies to various model calculations. 

Empirical extrapolation from lower energy data (named ``Busza'' in the figure)~\cite{Busza:2007ke} 
significantly under predicts the measurement at LHC energies. A simple power-law growth of 
charged-particle multiplicities near midrapidity in central Au+Au collisions seems to 
be followed up to RHIC energies (named as ``Barshay et. al'', in the figure)~\cite{Barshay:2001wp}.
Perturbative QCD-inspired Monte Carlo event generators, the HIJING model 
without jet quenching~\cite{Deng:2010mv}, the Dual Parton Model~\cite{Bopp:2007sa}
(named ``DPMJET III'' in the figure), 
or the Ultra-relativistic Quantum Molecular Dynamics model~\cite{Mitrovski:2008hb} 
(named ``UrQMD'' in the figure) are consistent with the measurement. The HIJING model
results without jet quenching were also consistent with the RHIC measurements. The
semi-microscopic models like LEXUS are successful in explaining the observed multiplicity
at RHIC (named as ``Jeon et. al'' in the figure)~\cite{Jeon:2000mc}.
Models based on initial-state gluon density saturation have a range of predictions 
depending on the specifics of the implementation~\cite{Albacete:2010fs,Levin:2010zy,Kharzeev:2004if,Kharzeev:2007zt,Armesto:2004ud}.
The best agreement with LHC data happens for model as described in (named as ``Kharzeev1'' and ``Armesto'' in 
the figure)~\cite{Kharzeev:2004if,Armesto:2004ud}. Similar are the conclusions for RHIC energy from these models.
The prediction of a hybrid model based on hydrodynamics and saturation of final-state phase space of scattered 
partons (named as ``Eskola'' in the figure)~\cite{Eskola:2001bf} is slightly on higher side compared to the measurement at LHC. 
But such a model seems to do a reasonable job for RHIC energies~\cite{Eskola:2001rx}.
Another hydrodynamic model in which multiplicity is scaled from p+p collisions over predicts
the measurement (named as ``Bozek'' in the
figure)~\cite{Bozek:2010wt}. 
Models incorporating constituent quark scaling and Landau
hydrodynamics  
(named as ``Sarkisyan'' in the figure)~\cite{Sarkisyan:2010kb} and based on modified PYTHIA and 
hadronic re-scattering (named as ``Humanic'' in the figure)~\cite{Humanic:2010su}, under predict the measurement at LHC energy.
At RHIC energies, models considering minijet production in ultra-relativistic heavy-ion collisions by 
taking semi-hard parton re-scatterings explicitly into account, under predict the multiplicities (named as ``Accardi et. al in
the figure)~\cite{Accardi:2001vu}. It is also seen at RHIC energies that models based on string fusion~\cite{Armesto:2001et}, dual string model~\cite{DiasdeDeus:2000cg}
seem to work well, whereas those based on heavy ion cascade LUCIFER model~\cite{Kahana:2001ky} under predicts the data.


  \begin{figure}[htbp]
\includegraphics[scale=0.4]{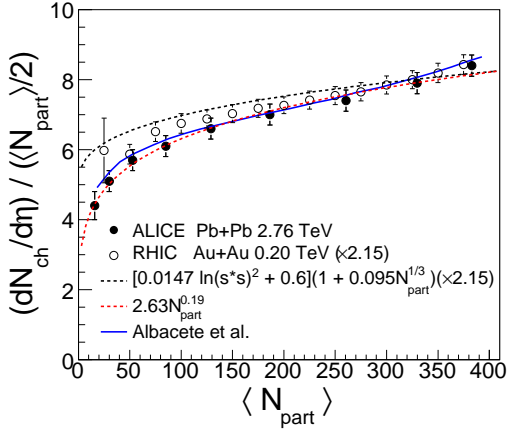}
    \caption{\label{fig:fig15}
      (Color online)  Centrality dependence of $(dN_{\rm
        {ch}}/d\eta)/(\langle N_{\rm {part}} \rangle/2)$ 
for Pb+Pb collisions at $\sqrt{s_{\rm {NN}}}$~=~2.76 TeV~\cite{Aamodt:2010cz} and 
Au+Au collisions at  $\sqrt{s_{\rm {NN}}}$~=~200~GeV. The RHIC results are scaled up by a factor of 2.15.
Also shown are comparisons to theoretical model calculations~\cite{ALbacete:2010ad} and some parametrization based on detail shape of
$dN_{\rm {ch}}/d\eta$ distributions at RHIC~\cite{Alver:2010ck} and $\langle N_{\rm {part}} \rangle$.  }
  \end{figure}

Figure~\ref{fig:fig15} shows the $(dN_{\rm {ch}}/d\eta)/(\langle N_{\rm {part}} \rangle/2)$ versus $\langle N_{\rm {part}} \rangle$
for Pb+Pb collisions at $\sqrt{s_{\rm {NN}}}$ = 2.76 TeV~\cite{Aamodt:2010cz}. Also shown are the corresponding RHIC results
scaled up by a factor 2.15. Remarkable similarity is observed in the shape of the distributions at RHIC and LHC energies.
Particle production based on saturation model explains the trends
nicely (named as ``Albacete et al'' in the
figure)~\cite{ALbacete:2010ad} (published after the most central dNch/deta value~\cite{Aamodt:2010pa} was known).
Simple fit to the data using a power law form for the $\langle N_{\rm {part}} \rangle$ also explains the measurements. 
In addition, a functional form inspired by the detailed shape of pseudorapidity distribution of charged particle multiplicity distributions
at RHIC~\cite{Alver:2010ck} explains the centrality trends nicely.

  \begin{figure}[htbp]
\includegraphics[scale=0.45]{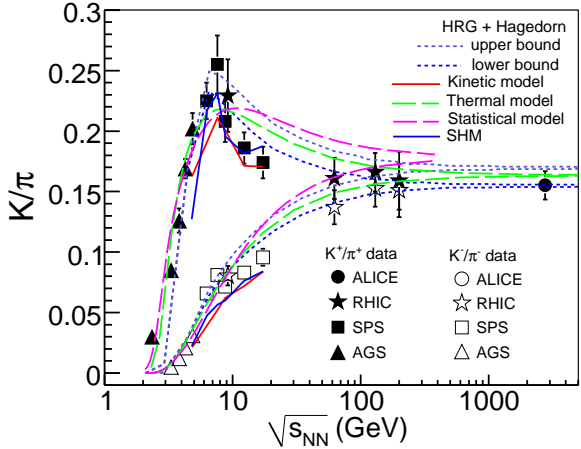}
    \caption{\label{fig:fig16}
      (Color online)    Energy dependence of $K^{\pm}/\pi^{\pm}$  ratio for central collisions at midrapidity. 
 Errors are statistical and systematic added in quadrature. Results are also compared with various
 theoretical model predictions~\cite{shm,stat,therm,kin,hrghag}.}
  \end{figure}

Strangeness production in heavy-ion collisions is a classic signature for formation of QGP~\cite{Koch:1986ud}.
The particle yield ratio $K/\pi$ could reflect the strangeness enhancement in heavy-ion collisions 
with respect to the elementary collisions.  Figure~\ref{fig:fig16} shows the energy dependence 
of $K^{\pm}/\pi^{\pm}$ ratio for central collisions at midrapidity. It will be interesting to
see which model explains such an impressive collection of systematic data on $K/\pi$ ratio.
Figure~\ref{fig:fig16} also shows the energy dependence of
$K/\pi$ ratio from various theoretical model calculations. 
The energy dependence of $K^{+}/\pi^{+}$ ratio has been
interpreted using the Statistical Model of Early Stage
(SMES)~\cite{smes}. The model predicts first order phase transition
and the existence of mixed phase around beam energy of 7-8 GeV.
The SHM or Statistical Hadronization Model~\cite{shm} assumes that the strong
interactions saturate the particle production matrix elements. 
This means that the yield of particles are controlled predominantly by the
magnitude of the accessible phase space. 
The system is in chemical non-equilibrium for 
$\sqrt{s_{NN}}<$ 7.6 GeV, while for higher energies, 
the over-saturation of chemical occupancies is observed.
  The Statistical Model~\cite{stat} assumes that the ratio
  of entropy to $T^3$ as a function of collision energy increases for
  mesons and decreases for baryons. Thus, a rapid change is expected
  at the crossing of the two curves, as the hadronic gas undergoes a
  transition from a baryon-dominated to a meson-dominated gas. The
  transition point is characterized by $T$=140 MeV, $\mu_B$= 410 MeV,
  and $\sqrt{s_{NN}}$=8.2 GeV. 
In the Thermal Model~\cite{therm}, the energy
  dependence of $K^{\pm}/\pi^{\pm}$ is studied by including
$\sigma$-meson, which is neglected in most
  of the models, and many higher
  mass resonances ($m>$ 2 GeV/$c^2$) into the resonance spectrum employed in
  the statistical model calculations. 
  The hadronic non-equilibrium Kinetic model~\cite{kin} assumes
  that the surplus of strange particles is produced in secondary
  reactions of hadrons generated in nuclear collisions. Then the two
  important aspects are the available energy density and the lifetime
  of the fireball.
It is suggested that these two aspects combine in such a way so as to show a sharp peak for
the strangeness-to-entropy or $K/\pi$ ratio as a function of beam energy.
In the Hadron Resonance Gas and Hagedorn model (HRG+Hagedorn)~\cite{hrghag}, all hadrons
  as given in PDG  with masses up to 2 GeV/$c^2$ are included. The unknown
  hadron resonances in this model are included through Hagedorn's
  formula for the density of states. The model assumes that the strangeness in the baryon sector
  decays to strange baryons and does not contribute to the kaon
  production.
The energy dependence of $K^{\pm}/\pi^{\pm}$ ratio
seems to be best explained using HRG+Hagedorn model.

This systematic measurement of $K/\pi$ ratio reveals two interesting information:
(a) The $K^{+}/\pi^{+}$ ratio shows a peak around $\sqrt{s_{\rm
    {NN}}}$ = 8 GeV, while the  $K^{-}/\pi^{-}$ ratio increases monotonically. The peak indicates the role of the maximum baryon density 
at freeze-out around this collision energy, and (b) for $\sqrt{s_{\rm {NN}}}$ $>$ 100 GeV, pair production 
becomes the dominant mechanism for  $K^{\pm}$ production, so both the ratios  $K^{+}/\pi^{+}$ and $K^{-}/\pi^{-}$
approach the value of 0.16. Taking into account of different masses between pions and kaons, this asymptotic  
value corresponding to a temperature of the order of 160 MeV.

\subsection{Azimuthal anisotropy}
  \begin{figure}[htbp]
\includegraphics[scale=0.42]{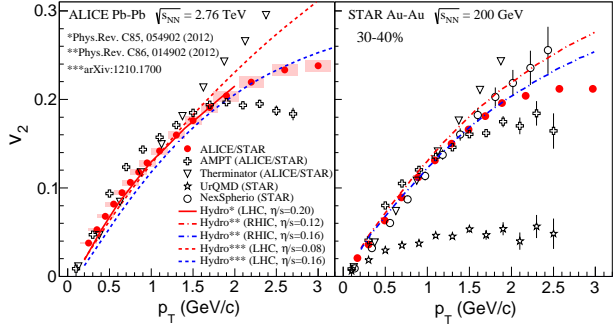}
    \caption{\label{fig:fig17}
      (Color online) The azimuthal anisotropy parameter $v_{2}$, measured in non-central heavy-ion collisions at midrapidity for RHIC and LHC energies.
For comparison, shown are the various theoretical calculations based on hydrodynamic and transport approaches (see text for details).  }
  \end{figure}
 The azimuthal anisotropy parameter $v_{2}$, measured at RHIC and LHC, provides an unique opportunity to
study the transport properties of the fundamental constituents of any visible matter - a system of
quarks and gluons. Furthermore, it provides an opportunity to understand whether the underlying dynamics of the
evolution of the system formed in the collisions is governed by macroscopic hydrodynamics~\cite{Shen:2012vn,Roy:2012jb,Roy:2012pn} 
or by microscopic transport approach~\cite{Xu:2011fi}. Figure~\ref{fig:fig17} shows the $v_{2}$ versus $p_{T}$ for 30-40\% collision centrality
Au+Au and Pb+Pb collisions at midrapidity for $\sqrt{s_{\rm {NN}}}$ = 200 GeV and 2.76 TeV, respectively.
The measurements are compared to a set of model calculations based on hydrodynamic approach 
(including THERMINATOR~\cite{Chojnacki:2011hb,Kisiel:2005hn}) and another
set of calculations based on transport approach. It is observed that hydrodynamic based models explain
the $v_{2}$ measurements both at RHIC and LHC energies. Transport based models including partonic interactions (like AMPT~\cite{Xu:2011fi})
also explain the $v_{2}$ measurements. However, those transport models which do not incorporate partonic
interactions like UrQMD~\cite{Bass:1998ca,Bleicher:1999xi} fail to
explain the data. The model comparison also reveals that the data favors a high degree of fluidity reflected by a small value of shear viscosity to entropy density ratio 
($\eta/s$) $<$ 0.2.
A more detailed comparison of the model calculations
with various order  azimuthal anisotropy parameters  $v_{n}$ would in near future give us a more quantitative
picture of the temperature (or energy) dependence of transport coefficients of the system formed in the
heavy-ion collisions.

\subsection{Nuclear modification factor}

\begin{widetext}

  \begin{figure}[htbp]
\includegraphics[scale=0.8]{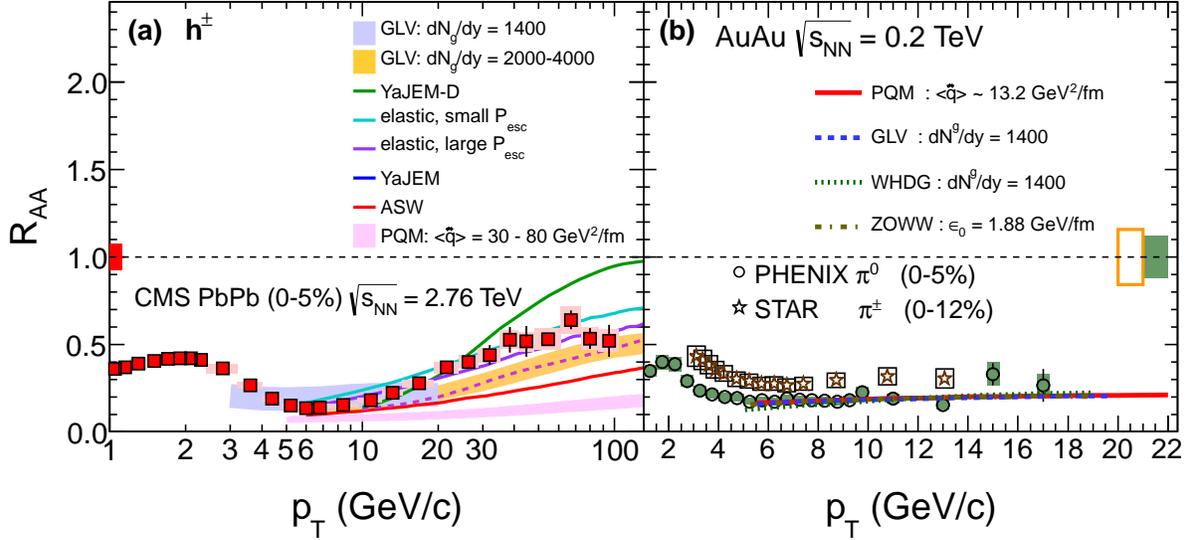}
\caption{\label{fig:fig18}
      (Color online)  Measurements of the nuclear modification factor $R_{\rm {AA}}$ in central heavy-ion collisions
at two different center-of-mass energies, as a function of $p_T$, for pions ($\pi^{\pm,0}$)~\cite{Adare:2008cg,Abelev:2007ra} and 
charged hadrons~\cite{Aamodt:2010jd,CMS:2012aa}, compared to several theoretical predictions (see text). The error bars on the points are the 
statistical uncertainties, and the boxes around the data points are the systematic uncertainties. 
Additional absolute normalization uncertainties of order  5\% to 10\% are not plotted. 
The bands for several of the theoretical calculations represent their uncertainties. }
  \end{figure}

The nuclear modification factor ($R_{\rm {AA}}$) is an observable used to study the
structure of strongly interacting dense matter formed in heavy-ion collisions. Here we discuss the observation
of $R_{\rm {AA}}$ $<$ 1 at high $p_{T}$ seen at RHIC and LHC by comparing to models within 
perturbative QCD (pQCD) based formalisms.
In this picture, the high $p_{T}$ hadrons are expected to originate from the fragmentation of hard partons
(hard scattering scales large than QCD scales of 200 MeV). The hard partons lose energy through interactions 
with the hot and dense medium, which get reflected in the observed values of $R_{\rm {AA}}$. The processes
by which they could loose energy includes radiative energy loss and elastic energy loss. For a more elaborate
discussion on these models we refer the reader to the review article~\cite{Majumder:2010qh}. 

In Fig.~\ref{fig:fig18}, we show a comparison between experimentally measured  $R_{\rm {AA}}$ versus  $p_{T}$ at LHC and RHIC energies
to corresponding pQCD based model calculations. All theoretical formalisms requires a microscopic model of 
the medium to set the input parameters for the energy loss calculation. These parameters for example are denoted as 
$\langle \hat{q} \rangle$, the transport coefficient of the medium or the gluon number density $dN^{g}/dy$ 
per unit rapidity. The parameter $P_{esc}$ on the other hand, reflects the strength of elastic energy loss 
put in the model calculations.  Without going into deeper theoretical 
discussions of each model, we refer the readers to the following related publications: PQM~\cite{Dainese:2004te}, GLV~\cite{Vitev:2002pf},
ASW~\cite{Armesto:2005iq}, YaJEM~\cite{Renk:2011gj}, WHDG~\cite{Wicks:2005gt} and ZOWW~\cite{Zhang:2007ja}. 
However, for completeness and to elucidate the approach taken in the model calculations, 
we briefly mention two formalisms as examples: the GLV approach named after their authors Gyulassy, Levai and Vitev 
and ASW approach named after the corresponding authors  Armesto,
Salgado and Wiedemann model, where the medium is defined as 
separated heavy static scattering centers with color screened potentials. Where as in some other formalism
a more precise definition of the medium is considered like being composed of quark gluon quasi-particles with dispersion 
relations and interactions given by the hard thermal loop effective theory. 

We observe that most models predict the $p_{T}$ dependence of $R_{\rm {AA}}$ well for collisions both at RHIC and LHC 
energies.  The models specially capture the generally rising behavior of $R_{\rm {AA}}$  
that is observed in the data at high $p_{T}$ for the LHC energies. The magnitude of the predicted slope of $R_{\rm {AA}}$ versus $p_{T}$ 
varies between models, depending on the assumptions for the jet-quenching mechanism. The models shown do not need 
larger values of medium density in the calculation to explain the $R_{\rm {AA}}$ for 3 $<$ $p_{\rm T}$ $<$ 20 GeV/$c$ 
at RHIC and LHC for the common kinematic range. They however require a high medium density at LHC energy to explain the 
values of $R_{\rm {AA}}$ for $p_{T}$ $>$ 20 GeV/$c$.

\end{widetext}

\section{Summary}
\label{sec:summary}

In summary, the results on multiplicity density in pseudorapidity,
HBT, azimuthal anisotropy, and nuclear modification factor from LHC experiments 
indicate that the fireball produced in these nuclear collisions is hotter, lives longer and expands to a larger size at freeze-out
compared to lower energies. These results also confirm the formation of a de-confined state of quarks and gluons at RHIC energies. 
The measurements at LHC provide a unique kinematic access to study in detail, the properties (such as transport co-efficients) 
of this system of quarks and gluons. 

In this review, we showed that the first set of  measurements made by the three LHC experiments within the 
heavy-ion program, ALICE, ATLAS and CMS, show a high degree of consistency. These measurements includes,
centrality dependence of charged particle multiplicity, azimuthal anisotropy and nuclear modification
factor versus transverse momentum. Next, we discussed the comparison of various measurements
made at RHIC and LHC energies. LHC measurements of   $dN_{\rm {ch}}/d\eta$ clearly demonstrated the power law
dependence of charged particle multiplicity on the beam energy. They also reconfirmed the observation at RHIC
that particle production mechanism is not a simple superposition of several $p$+$p$ collisions. The values of $\langle m_{T} \rangle$,
$\epsilon_{\rm {Bj}}$, freeze-out volume, decoupling time for hadrons, $\langle v_{2} \rangle$ and $\langle \beta \rangle$ 
are larger at LHC energies compared to those at RHIC. This is even though the freeze-out temperatures are comparable. The value of net-charge
fluctuation measure is observed to rapidly approach towards a simple model based calculation for QGP state. 
However, the sensitivity of this observable for a heavy-ion system as well as lack of proper modeling of the heavy-ion
system theoretically for such an observable needs careful consideration. The 
$v_{2}$ fluctuations as a function of centrality fraction have a similar value at both RHIC and LHC. This reflects their
sensitivity to initial state effects. 
Just like at RHIC the $R_{\rm {dAu}}$ and direct photon $R_{\rm {AA}}$ measurements experimentally 
demonstrated that the observed  $R_{\rm {AA}}$ $<$ 1 for charged hadrons is a final state effect, so also at LHC the $R_{\rm {pPb}}$, 
direct photon, $W^{\pm}$ and $Z^{0}$ $R_{\rm {AA}}$ measurements showed that the observed  $R_{\rm {AA}}$ $<$ 1  
is indeed due to formation of a dense medium of colored charges in central heavy-ion collisions.
All these conclusions were further validated by the comparison of several observables to corresponding model calculations. 
Further it was found that the fluid at LHC shows a comparable degree of fluidity as that at RHIC. This is reflected
by a small value of shear viscosity to entropy density ratio. 

Measurements related heavy quark production~\cite{Chatrchyan:2012np,ALICE:2013xna,Adamczyk:2012pw}, 
dilepton production, jet-hadron correlations~\cite{Adamczyk:2013jei,CMS:2012qk}
and higher order azimuthal anisotropy~\cite{:2013waa,Aad:2013fja} 
which are now coming out of both RHIC and LHC experiments will provide a much more detail characterization of the properties 
of the QCD matter formed in heavy-ion collisions.

\noindent{\bf Acknowledgments}\\
We would like to thank F. Antinori, S. Gupta, D. Keane, A. K. Mohanty, Y. P. Viyogi, and N. Xu, for reading the manuscript,
helpful discussions and comments. BM is supported by the DAE-SRC project fellowship for this work. LK is partly supported
by DOE grant DE-FG02-89ER40531 for carrying out this work.


\normalsize
\begin{thebibliography}{99}


\bibitem{Mohanty:2011ki} 
  B.~Mohanty,
  New J.\ Phys.\  {\bf 13}, 065031 (2011)
  [arXiv:1102.2495 [nucl-ex]].

\bibitem{Mohanty:2009vb} 
  B.~Mohanty,
  Nucl.\ Phys.\ A {\bf 830}, 899C (2009)
  [arXiv:0907.4476 [nucl-ex]].

\bibitem{Fukushima:2010bq} 
  K.~Fukushima and T.~Hatsuda,
  Rept.\ Prog.\ Phys.\  {\bf 74}, 014001 (2011)
  [arXiv:1005.4814 [hep-ph]].



\bibitem{Gupta:2011wh}
  S.~Gupta, X.~Luo, B.~Mohanty, H.~G.~Ritter and N.~Xu,
  Science {\bf 332} (2011) 1525
  [arXiv:1105.3934 [hep-ph]].


\bibitem{Bazavov:2011nk} 
  A.~Bazavov, T.~Bhattacharya, M.~Cheng, C.~DeTar, H.~T.~Ding, S.~Gottlieb, R.~Gupta and P.~Hegde {\it et al.},
  Phys.\ Rev.\ D {\bf 85}, 054503 (2012)
  [arXiv:1111.1710 [hep-lat]].

\bibitem{Borsanyi:2010bp} 
  S.~Borsanyi {\it et al.}  [Wuppertal-Budapest Collaboration],
  JHEP {\bf 1009}, 073 (2010)
  [arXiv:1005.3508 [hep-lat]].





\bibitem{Arsene:2004fa} 
  I.~Arsene {\it et al.}  [BRAHMS Collaboration],
  Nucl.\ Phys.\ A {\bf 757}, 1 (2005)
  [nucl-ex/0410020].

\bibitem{Back:2004je} 
  B.~B.~Back, M.~D.~Baker, M.~Ballintijn, D.~S.~Barton, B.~Becker, R.~R.~Betts, A.~A.~Bickley and R.~Bindel {\it et al.},
  Nucl.\ Phys.\ A {\bf 757}, 28 (2005)
  [nucl-ex/0410022].



\bibitem{Adams:2005dq} 
  J.~Adams {\it et al.}  [STAR Collaboration],
  Nucl.\ Phys.\ A {\bf 757}, 102 (2005)
  [nucl-ex/0501009].

\bibitem{Adcox:2004mh} 
  K.~Adcox {\it et al.}  [PHENIX Collaboration],
  Nucl.\ Phys.\ A {\bf 757}, 184 (2005)
  [nucl-ex/0410003].


\bibitem{Gyulassy:2004zy} 
  M.~Gyulassy and L.~McLerran,
  Nucl.\ Phys.\ A {\bf 750}, 30 (2005)
  [nucl-th/0405013].


\bibitem{Mohanty:2011nm} 
  B.~Mohanty [STAR Collaboration],
  J.\ Phys.\ G {\bf 38}, 124023 (2011)
  [arXiv:1106.5902 [nucl-ex]].


\bibitem{Miller:2007ri} 
  M.~L.~Miller, K.~Reygers, S.~J.~Sanders and P.~Steinberg,
  Ann.\ Rev.\ Nucl.\ Part.\ Sci.\  {\bf 57}, 205 (2007)
  [nucl-ex/0701025].




\bibitem{Aamodt:2010cz} 
  K.~Aamodt {\it et al.}  [ALICE Collaboration],
  Phys.\ Rev.\ Lett.\  {\bf 106}, 032301 (2011)
  [arXiv:1012.1657 [nucl-ex]].

\bibitem{Chatrchyan:2011pb} 
  S.~Chatrchyan {\it et al.}  [CMS Collaboration],
  JHEP {\bf 1108}, 141 (2011)
  [arXiv:1107.4800 [nucl-ex]];



\bibitem{ATLAS:2011ag} 
  G.~Aad {\it et al.}  [ATLAS Collaboration],
  Phys.\ Lett.\ B {\bf 710}, 363 (2012)
  [arXiv:1108.6027 [hep-ex]];

\bibitem{Chatrchyan:2012xq} 
  S.~Chatrchyan {\it et al.}  [CMS Collaboration],
  Phys.\ Rev.\ Lett.\  {\bf 109}, 022301 (2012)
  [arXiv:1204.1850 [nucl-ex]];
ibid.,
    Phys. Rev. Lett. {\bf 110}, 042301 (2013); 
ibid.,
Phys. Rev. C {\bf 87}, 014902 (2013);
ibid.,
Eur. Phys. J. C {\bf 72}, 10052 (2012).


\bibitem{ATLAS:2011ah} 
  G.~Aad {\it et al.}  [ATLAS Collaboration],
  Phys.\ Lett.\ B {\bf 707}, 330 (2012)
  [arXiv:1108.6018 [hep-ex]];
ibid.,
arXiv:1305.2942 [hep-ex];
ibid.,
Phys. Rev. C  {\bf 86}, 014907 (2012);
J. Jia (for the ATLAS Collaboration), arXiv:1208.1427 [nucl-ex].


\bibitem{Aamodt:2010pa} 
  K.~Aamodt {\it et al.}  [ALICE Collaboration],
  Phys.\ Rev.\ Lett.\  {\bf 105}, 252302 (2010)
  [arXiv:1011.3914 [nucl-ex]].



\bibitem{Aamodt:2010jd} 
  K.~Aamodt {\it et al.}  [ALICE Collaboration],
  Phys.\ Lett.\ B {\bf 696}, 30 (2011)
  [arXiv:1012.1004 [nucl-ex]].


\bibitem{CMS:2012aa} 
  S.~Chatrchyan {\it et al.}  [CMS Collaboration],
  Eur.\ Phys.\ J.\ C {\bf 72}, 1945 (2012)
  [arXiv:1202.2554 [nucl-ex]].

\bibitem{Back:2003hx} 
  B.~B.~Back {\it et al.}  [PHOBOS Collaboration],
  Phys.\ Rev.\ Lett.\  {\bf 93}, 082301 (2004)
  [nucl-ex/0311009].


\bibitem{Abreu:2002fx} 
  M.~C.~Abreu {\it et al.}  [NA50 Collaboration],
  Phys.\ Lett.\ B {\bf 530}, 33 (2002).

\bibitem{Adler:2001yq} 
  C.~Adler {\it et al.}  [STAR Collaboration],
  Phys.\ Rev.\ Lett.\  {\bf 87}, 112303 (2001)
  [nucl-ex/0106004].



\bibitem{Bearden:2001xw} 
  I.~G.~Bearden {\it et al.}  [BRAHMS Collaboration],
  Phys.\ Lett.\ B {\bf 523}, 227 (2001)
  [nucl-ex/0108016].



\bibitem{Bearden:2001qq} 
  I.~G.~Bearden {\it et al.}  [BRAHMS Collaboration],
  Phys.\ Rev.\ Lett.\  {\bf 88}, 202301 (2002)
  [nucl-ex/0112001].



\bibitem{Adcox:2000sp} 
  K.~Adcox {\it et al.}  [PHENIX Collaboration],
  Phys.\ Rev.\ Lett.\  {\bf 86}, 3500 (2001)
  [nucl-ex/0012008].



\bibitem{Back:2000gw} 
  B.~B.~Back {\it et al.}  [PHOBOS Collaboration],
  Phys.\ Rev.\ Lett.\  {\bf 85}, 3100 (2000)
  [hep-ex/0007036].



\bibitem{Back:2002wb} 
  B.~B.~Back, M.~D.~Baker, D.~S.~Barton, R.~R.~Betts, M.~Ballintijn, A.~A.~Bickley, R.~Bindel and A.~Budzanowski {\it et al.},
  Phys.\ Rev.\ Lett.\  {\bf 91}, 052303 (2003)
  [nucl-ex/0210015].



\bibitem{Aamodt:2010pb} 
  KAamodt {\it et al.}  [ALICE Collaboration],
  Phys.\ Rev.\ Lett.\  {\bf 105}, 252301 (2010)
  [arXiv:1011.3916 [nucl-ex]].


\bibitem{ua1}
 C. Albajar {\it et al.}  [UA1 Collaboration], Nucl.\ Phys.\ B {\bf 335}, 261 (1990). 

\bibitem{ua5}
 K. Alpgard {\it et al.}  [UA5 Collaboration], Phys.\ Lett.\ B {\bf 121}, 209 (1983); 
 G. Alner   {\it et al.}  [UA5 Collaboration], Z. \ Phys.\ C {\bf 33}, 1 (1986); 


\bibitem{Abelev:2008ab} 
  B.~I.~Abelev {\it et al.}  [STAR Collaboration],
  Phys.\ Rev.\ C {\bf 79}, 034909 (2009)
  [arXiv:0808.2041 [nucl-ex]].


\bibitem{Abe:1989td} 
  F.~Abe {\it et al.}  [CDF Collaboration],
  Phys.\ Rev.\ D {\bf 41}, 2330 (1990).

\bibitem{Aamodt:2010pp} 
  K.~Aamodt {\it et al.}  [ALICE Collaboration],
  Eur.\ Phys.\ J.\ C {\bf 68}, 345 (2010)
  [arXiv:1004.3514 [hep-ex]].



\bibitem{Khachatryan:2010xs} 
  V.~Khachatryan {\it et al.}  [CMS Collaboration],
  JHEP {\bf 1002}, 041 (2010)
  [arXiv:1002.0621 [hep-ex]].

\bibitem{isr} 
 W. Thome {\it et al.}, Nucl.\ Phys.\ B {\bf 129}, 1 (1986).

\bibitem{Aamodt:2010ft} 
  K.~Aamodt {\it et al.}  [ALICE Collaboration],
  Eur.\ Phys.\ J.\ C {\bf 68}, 89 (2010)
  [arXiv:1004.3034 [hep-ex]].

\bibitem{Alver:2010ck} 
  B.~Alver {\it et al.}  [PHOBOS Collaboration],
  Phys.\ Rev.\ C {\bf 83}, 024913 (2011)
  [arXiv:1011.1940 [nucl-ex]].
 
\bibitem{ALICE:2012xs} 
  B.~Abelev {\it et al.}  [ALICE Collaboration],
  Phys.\  Rev.\  Lett.\  110, {\bf 032301} (2013)
  [Phys.\ Rev.\ Lett.\  {\bf 110}, 032301 (2013)]
  [arXiv:1210.3615 [nucl-ex]].









\bibitem{Alber:1997sn} 
  T.~Alber {\it et al.}  [NA35 Collaboration],
  Eur.\ Phys.\ J.\ C {\bf 2}, 643 (1998)
  [hep-ex/9711001].


\bibitem{mishra}
A.~N.~Mishra, R.~Mazumder and R.~Sahoo,
  arXiv:1304.2113 [hep-ph].


\bibitem{Ahle:1999uy} 
  L.~Ahle {\it et al.}  [E866 and E917 Collaborations],
  Phys.\ Lett.\ B {\bf 476}, 1 (2000)
  [nucl-ex/9910008].

\bibitem{Barrette:1999ry} 
  J.~Barrette {\it et al.}  [E877 Collaboration],
  Phys.\ Rev.\ C {\bf 62}, 024901 (2000)
  [nucl-ex/9910004].

\bibitem{Afanasiev:2002mx} 
  S.~V.~Afanasiev {\it et al.}  [NA49 Collaboration],
  Phys.\ Rev.\ C {\bf 66}, 054902 (2002)
  [nucl-ex/0205002].


\bibitem{Alt:2007aa} 
  C.~Alt {\it et al.}  [NA49 Collaboration],
  Phys.\ Rev.\ C {\bf 77}, 024903 (2008)
  [arXiv:0710.0118 [nucl-ex]].


\bibitem{Abelev:2009bw} 
  B.~I.~Abelev {\it et al.}  [STAR Collaboration],
  Phys.\ Rev.\ C {\bf 81}, 024911 (2010)
  [arXiv:0909.4131 [nucl-ex]].




\bibitem{Abelev:2012wca} 
  B.~Abelev {\it et al.}  [ALICE Collaboration],
  Phys.\ Rev.\ Lett.\  {\bf 109}, 252301 (2012)
  [arXiv:1208.1974 [hep-ex]].


\bibitem{Bjorken:1982qr} 
  J.~D.~Bjorken,
  Phys.\ Rev.\ D {\bf 27}, 140 (1983).


\bibitem{Loizides:2011ys} 
  C.~Loizides,
  J.\ Phys.\ G {\bf 38}, 124040 (2011)
  [arXiv:1106.6324 [nucl-ex]].


\bibitem{Chatrchyan:2012mb} 
  S.~Chatrchyan {\it et al.}  [CMS Collaboration],
  Phys.\ Rev.\ Lett.\  {\bf 109}, 152303 (2012)
  [arXiv:1205.2488 [nucl-ex]].

\bibitem{Mitchell:2012mx} 
  J.~T.~Mitchell [PHENIX Collaboration],
  arXiv:1211.6139 [nucl-ex].


\bibitem{Adler:2004zn} 
  S.~S.~Adler {\it et al.}  [PHENIX Collaboration],
  Phys.\ Rev.\ C {\bf 71}, 034908 (2005)
  [Erratum-ibid.\ C {\bf 71}, 049901 (2005)]
  [nucl-ex/0409015].

\bibitem{Adcox:2001ry} 
  K.~Adcox {\it et al.}  [PHENIX Collaboration],
  Phys.\ Rev.\ Lett.\  {\bf 87}, 052301 (2001)
  [nucl-ex/0104015].

\bibitem{Aggarwal:2000bc} 
  M.~M.~Aggarwal {\it et al.}  [WA98 Collaboration],
  Eur.\ Phys.\ J.\ C {\bf 18}, 651 (2001)
  [nucl-ex/0008004].


\bibitem{Mohanty:2003fy} 
  B.~Mohanty, J.~-eAlam, S.~Sarkar, T.~K.~Nayak, B.~KNandi and ,
  Phys.\ Rev.\ C {\bf 68}, 021901 (2003)
  [nucl-th/0304023].

\bibitem{vanhove} L.~Van Hove, Phys.\ Lett. \ B {\bf 118}, 138 (1982).




\bibitem{Lisa:2000hw} 
  M.~A.~Lisa {\it et al.}  [E895 Collaboration],
  Phys.\ Rev.\ Lett.\  {\bf 84}, 2798 (2000).

\bibitem{Alt:2007uj} 
  C.~Alt {\it et al.}  [NA49 Collaboration],
  Phys.\ Rev.\ C {\bf 77}, 064908 (2008)
  [arXiv:0709.4507 [nucl-ex]].

\bibitem{Adamova:2002wi} 
  D.~Adamova {\it et al.}  [CERES Collaboration],
  Nucl.\ Phys.\ A {\bf 714}, 124 (2003)
  [nucl-ex/0207005].


\bibitem{Abelev:2009tp} 
  B.~I.~Abelev {\it et al.}  [STAR Collaboration],
  Phys.\ Rev.\ C {\bf 80}, 024905 (2009)
  [arXiv:0903.1296 [nucl-ex]].

\bibitem{Back:2004ug} 
  B.~B.~Back {\it et al.}  [PHOBOS Collaboration],
  Phys.\ Rev.\ C {\bf 73}, 031901 (2006)
  [nucl-ex/0409001].

\bibitem{Aamodt:2011mr} 
  K.~Aamodt {\it et al.}  [ALICE Collaboration],
  Phys.\ Lett.\ B {\bf 696}, 328 (2011)
  [arXiv:1012.4035 [nucl-ex]].


\bibitem{Bertsch:1989vn} 
  G.~F.~Bertsch,
  Nucl.\ Phys.\ A {\bf 498}, 173C (1989).

\bibitem{Pratt:1986cc} 
  S.~Pratt,
  Phys.\ Rev.\ D {\bf 33}, 1314 (1986).


\bibitem{Teaney:2003kp} 
  D.~Teaney,
  Phys.\ Rev.\ C {\bf 68}, 034913 (2003)
  [nucl-th/0301099].



\bibitem{Reisdorf:1996qj} 
  W.~Reisdorf {\it et al.}  [FOPI Collaboration],
  Nucl.\ Phys.\ A {\bf 612}, 493 (1997)
  [nucl-ex/9610009].



\bibitem{Lisa:1994yr} 
  M.~A.~Lisa {\it et al.}  [EOS Collaboration],
  Phys.\ Rev.\ Lett.\  {\bf 75}, 2662 (1995)
  [nucl-ex/9502001].

\bibitem{Muntz:1997qt} 
  C.~Muntz [E802 Collaboration],
  nucl-ex/9806002.

\bibitem{Appelshauser:1997rr} 
  H.~Appelshauser {\it et al.}  [NA49 Collaboration],
  Eur.\ Phys.\ J.\ C {\bf 2}, 661 (1998)
  [hep-ex/9711024].

\bibitem{Andronic:2005yp} 
  A.~Andronic, P.~Braun-Munzinger, and J.~Stachel,
  Nucl.\ Phys.\ A {\bf 772}, 167 (2006)
  [nucl-th/0511071].

\bibitem{Andronic:2008gu} 
  A.~Andronic, P.~Braun-Munzinger, and J.~Stachel,
  Phys.\ Lett.\ B {\bf 673}, 142 (2009)
  [Erratum-ibid.\ B {\bf 678}, 516 (2009)]
  [arXiv:0812.1186 [nucl-th]].


\bibitem{Cleymans:1998yf} 
  J.~Cleymans, H.~Oeschler, K.~Redlich and ,
  J.\ Phys.\ G {\bf 25}, 281 (1999)
  [nucl-th/9809031].

\bibitem{Cleymans:1997sw} 
  J.~Cleymans, D.~Elliott, A.~Keranen, E.~Suhonen and ,
  Phys.\ Rev.\ C {\bf 57}, 3319 (1998)
  [nucl-th/9711066].


\bibitem{BraunMunzinger:1994xr} 
  P.~Braun-Munzinger, J.~Stachel, J.~P.~Wessels, N.~Xu and ,
  Phys.\ Lett.\ B {\bf 344}, 43 (1995)
  [nucl-th/9410026].

\bibitem{BraunMunzinger:1999qy} 
  P.~Braun-Munzinger, I.~Heppe, J.~Stachel and ,
  Phys.\ Lett.\ B {\bf 465}, 15 (1999)
  [nucl-th/9903010].

\bibitem{Becattini:2002ix} 
  F.~Becattini, G.~Pettini and ,
  Phys.\ Rev.\ C {\bf 67}, 015205 (2003)
  [hep-ph/0204340].

\bibitem{Becattini:2000jw} 
  F.~Becattini, J.~Cleymans, A.~Keranen, ESuhonen, K.~Redlich and ,
  Phys.\ Rev.\ C {\bf 64}, 024901 (2001)
  [hep-ph/0002267].




\bibitem{Milano:2013sza} 
  L.~Milano [ for the ALICE Collaboration],
  arXiv:1302.6624 [hep-ex].

\bibitem{Steinheimer:2012rd} 
  J.~Steinheimer, J.~Aichelin, M.~Bleicher and ,
  Phys.\ Rev.\ Lett.\  {\bf 110}, 042501 (2013)
  [arXiv:1203.5302 [nucl-th]].



\bibitem{Sako:2004pw} 
  H.~Sako {\it et al.}  [CERES/NA45 Collaboration],
  J.\ Phys.\ G {\bf 30}, S1371 (2004)
  [nucl-ex/0403037].


\bibitem{Abelev:2008jg} 
  B.~I.~Abelev {\it et al.}  [STAR Collaboration],
  Phys.\ Rev.\ C {\bf 79}, 024906 (2009)
  [arXiv:0807.3269 [nucl-ex]].

\bibitem{Abelev:2012pv} 
  B.~Abelev {\it et al.}  [ALICE Collaboration],
  arXiv:1207.6068 [nucl-ex].

\bibitem{Jeon:2000wg} 
  S.~Jeon, V.~Koch and ,
  Phys.\ Rev.\ Lett.\  {\bf 85}, 2076 (2000)
  [hep-ph/0003168].


\bibitem{Pruneau:2002yf} 
  C.~Pruneau, S.~Gavin, S.~Voloshin and ,
  Phys.\ Rev.\ C {\bf 66}, 044904 (2002)
  [nucl-ex/0204011].



\bibitem{Ollitrault:1992bk} 
  J.~-Y.~Ollitrault,
  Phys.\ Rev.\ D {\bf 46}, 229 (1992).


\bibitem{Andronic:2004cp} 
  A.~Andronic {\it et al.}  [FOPI Collaboration],
  Phys.\ Lett.\ B {\bf 612}, 173 (2005)
  [nucl-ex/0411024].

\bibitem{Bastid:2005ct} 
  N.~Bastid {\it et al.}  [FOPI Collaboration],
  Phys.\ Rev.\ C {\bf 72}, 011901 (2005)
  [nucl-ex/0504002].

\bibitem{BraunMunzinger:1998cg} 
  P.~Braun-Munzinger, J.~Stachel and ,
  Nucl.\ Phys.\ A {\bf 638}, 3 (1998)
  [nucl-ex/9803015].

\bibitem{Pinkenburg:1999ya} 
  C.~Pinkenburg {\it et al.}  [E895 Collaboration],
  Phys.\ Rev.\ Lett.\  {\bf 83}, 1295 (1999)
  [nucl-ex/9903010].

\bibitem{Adamova:2002qx} 
  D.~Adamova {\it et al.}  [CERES Collaboration],
  Nucl.\ Phys.\ A {\bf 698}, 253 (2002).

\bibitem{Alt:2003ab} 
  C.~Alt {\it et al.}  [NA49 Collaboration],
  Phys.\ Rev.\ C {\bf 68}, 034903 (2003)
  [nucl-ex/0303001].

\bibitem{Adams:2004bi} 
  J.~Adams {\it et al.}  [STAR Collaboration],
  Phys.\ Rev.\ C {\bf 72}, 014904 (2005)
  [nucl-ex/0409033].

\bibitem{Back:2004zg} 
  B.~B.~Back {\it et al.}  [PHOBOS Collaboration],
  Phys.\ Rev.\ Lett.\  {\bf 94}, 122303 (2005)
  [nucl-ex/0406021].

\bibitem{Afanasiev:2009wq} 
  S.~Afanasiev {\it et al.}  [PHENIX Collaboration],
  Phys.\ Rev.\ C {\bf 80}, 024909 (2009)
  [arXiv:0905.1070 [nucl-ex]].

\bibitem{ATLAS:2012gna} 
  [ATLAS Collaboration],
  ATLAS-CONF-2012-117.


\bibitem{Adare:2011tg} 
  A.~Adare {\it et al.}  [PHENIX Collaboration],
  Phys.\ Rev.\ Lett.\  {\bf 107}, 252301 (2011)
  [arXiv:1105.3928 [nucl-ex]].

\bibitem{ALICE:2011ab} 
  K.~Aamodt {\it et al.}  [ALICE Collaboration],
  Phys.\ Rev.\ Lett.\  {\bf 107}, 032301 (2011)
  [arXiv:1105.3865 [nucl-ex]].


\bibitem{Adare:2012vq} 
  A.~Adare {\it et al.}  [PHENIX Collaboration],
  Phys.\ Rev.\ C {\bf 85}, 064914 (2012)
  [arXiv:1203.2644 [nucl-ex]].

\bibitem{Abelev:2012di} 
  B.~Abelev {\it et al.}  [ALICE Collaboration],
  arXiv:1205.5761 [nucl-ex].





\bibitem{Muller:2005pv} 
  B.~Muller, R.~J.~Fries and S.~A.~Bass,
  Phys.\ Lett.\ B {\bf 618}, 77 (2005)
  [nucl-th/0503003].

\bibitem{Greco:2005jk} 
  V.~Greco and C.~M.~Ko,
  nucl-th/0505061.





\bibitem{Agakishiev:2011eq} 
  G.~Agakishiev {\it et al.}  [STAR Collaboration],
  Phys.\ Rev.\ C {\bf 86}, 014904 (2012)
  [arXiv:1111.5637 [nucl-ex]].


\bibitem{Voloshin:2007pc} 
  S.~A.~Voloshin, A.~M.~Poskanzer, A.~Tang and G.~Wang,
  Phys.\ Lett.\ B {\bf 659}, 537 (2008)
  [arXiv:0708.0800 [nucl-th]].

\bibitem{Voloshin:1994mz} 
  S.~Voloshin and Y.~Zhang,
  Z.\ Phys.\ C {\bf 70}, 665 (1996)
  [hep-ph/9407282].


\bibitem{Voloshin:2008dg} 
  S.~A.~Voloshin, A.~M.~Poskanzer and R.~Snellings,
  arXiv:0809.2949 [nucl-ex].

\bibitem{Aad:2013xma} 
  G.~Aad {\it et al.}  [ATLAS Collaboration],
  arXiv:1305.2942 [hep-ex].




\bibitem{Adams:2003im} 
  J.~Adams {\it et al.}  [STAR Collaboration],
  Phys.\ Rev.\ Lett.\  {\bf 91}, 072304 (2003)
  [nucl-ex/0306024].

\cite{Adler:2006bw}
\bibitem{Adler:2006bw} 
  S.~S.~Adler {\it et al.}  [PHENIX Collaboration],
  Phys.\ Rev.\ C {\bf 76}, 034904 (2007)
  [nucl-ex/0611007].



\bibitem{ALICE:2012mj} 
  B.~Abelev {\it et al.}  [ALICE Collaboration],
  arXiv:1210.4520 [nucl-ex].


\bibitem{Antreasyan:1977aw} 
  D.~Antreasyan, J.~W.~Cronin, H.~J.~Frisch, M.~J.~Shochet, L.~Kluberg, P.~A.~Piroue and R.~L.~Sumner,
  Phys.\ Rev.\ Lett.\  {\bf 38}, 115 (1977).

\bibitem{Antreasyan:1977au} 
  D.~Antreasyan, J.~W.~Cronin, H.~J.~Frisch, M.~J.~Shochet, L.~Kluberg, P.~A.~Piroue and R.~L.~Sumner,
  Phys.\ Rev.\ Lett.\  {\bf 38}, 112 (1977).





\bibitem{Adler:2005ig} 
  S.~S.~Adler {\it et al.}  [PHENIX Collaboration],
  Phys.\ Rev.\ Lett.\  {\bf 94}, 232301 (2005)
  [nucl-ex/0503003].

\bibitem{Chatrchyan:2012vq} 
  S.~Chatrchyan {\it et al.}  [CMS Collaboration],
  Phys.\ Lett.\ B {\bf 710}, 256 (2012)
  [arXiv:1201.3093 [nucl-ex]].


\bibitem{Chatrchyan:2012nt} 
  S.~Chatrchyan {\it et al.}  [CMS Collaboration],
  Phys.\ Lett.\ B {\bf 715}, 66 (2012)
  [arXiv:1205.6334 [nucl-ex]].

\bibitem{Chatrchyan:2011ua} 
  S.~Chatrchyan {\it et al.}  [CMS Collaboration],
  Phys.\ Rev.\ Lett.\  {\bf 106}, 212301 (2011)
  [arXiv:1102.5435 [nucl-ex]].







\bibitem{Busza:2007ke} 
  W.~Busza,
  J.\ Phys.\ G {\bf 35}, 044040 (2008)
  [arXiv:0710.2293 [nucl-ex]].


\bibitem{Barshay:2001wp} 
  S.~Barshay, G.~Kreyerhoff and ,
  Nucl.\ Phys.\ A {\bf 697}, 563 (2002)
  [Erratum-ibid.\ A {\bf 703}, 891 (2002)]
  [hep-ph/0104303].


\bibitem{Deng:2010mv} 
  W.~-T.~Deng, X.~-N.~Wang, R.~Xu and ,
  Phys.\ Rev.\ C {\bf 83}, 014915 (2011)
  [arXiv:1008.1841 [hep-ph]].


\bibitem{Bopp:2007sa} 
  F.~W.~Bopp, R.~Engel, J.~Ranft, S.~Roesler and ,
  arXiv:0706.3875 [hep-ph].


\bibitem{Mitrovski:2008hb} 
  M.~Mitrovski, T.~Schuster, G.~Graf, H.~Petersen, M.~Bleicher and ,
  Phys.\ Rev.\ C {\bf 79}, 044901 (2009)
  [arXiv:0812.2041 [hep-ph]].


\bibitem{Jeon:2000mc} 
  S.~Jeon, J.~I.~Kapusta and ,
  Phys.\ Rev.\ C {\bf 63}, 011901 (2001)
  [nucl-th/0009032].


\bibitem{Albacete:2010fs} 
  J.~LAlbacete,
  J.\ Phys.\ Conf.\ Ser.\  {\bf 270}, 012052 (2011)
  [arXiv:1010.6027 [hep-ph]].

\bibitem{Levin:2010zy} 
  E.~Levin, A.~H.~Rezaeian and ,
  Phys.\ Rev.\ D {\bf 82}, 054003 (2010)
  [arXiv:1007.2430 [hep-ph]].


\bibitem{Kharzeev:2004if} 
  D.~Kharzeev, E.~Levin, M.~Nardi and ,
  Nucl.\ Phys.\ A {\bf 747}, 609 (2005)
  [hep-ph/0408050].

\bibitem{Kharzeev:2007zt} 
  D.~Kharzeev, E.~Levin, M.~Nardi and ,
  arXiv:0707.0811 [hep-ph].

\bibitem{Armesto:2004ud} 
  N.~Armesto, C.~A.~Salgado, U.~A.~Wiedemann and ,
  Phys.\ Rev.\ Lett.\  {\bf 94}, 022002 (2005)
  [hep-ph/0407018].


\bibitem{Eskola:2001bf} 
  K.~J.~Eskola, P.~V.~Ruuskanen, S.~S.~Rasanen, K.~Tuominen and ,
  Nucl.\ Phys.\ A {\bf 696}, 715 (2001)
  [hep-ph/0104010].


\bibitem{Eskola:2001rx} 
  K.~J.~Eskola, K.~Kajantie, K.~Tuominen and ,
  Nucl.\ Phys.\ A {\bf 700}, 509 (2002)
  [hep-ph/0106330].



\bibitem{Bozek:2010wt} 
  P.~Bozek, M.~Chojnacki, W.~Florkowski, B.~Tomasik and ,
  Phys.\ Lett.\ B {\bf 694}, 238 (2010)
  [arXiv:1007.2294 [nucl-th]].


\bibitem{Sarkisyan:2010kb} 
  E.~K.~G.~Sarkisyan, A.~S.~Sakharov and ,
  Eur.\ Phys.\ J.\ C {\bf 70}, 533 (2010)
  [arXiv:1004.4390 [hep-ph]];
  E.~K.~G.~Sarkisyan and A.~S.~Sakharov,
  AIP Conf.\ Proc.\  {\bf 828}, 35 (2006)
  [hep-ph/0510191].

\bibitem{Humanic:2010su} 
  T.~J.~Humanic,
  arXiv:1011.0378 [nucl-th].

\bibitem{Accardi:2001vu} 
  A.~Accardi,
  hep-ph/0104060.




\bibitem{Armesto:2001et} 
  N.~Armesto, C.~Pajares, D.~Sousa and ,
  Phys.\ Lett.\ B {\bf 527}, 92 (2002)
  [hep-ph/0104269].




\bibitem{DiasdeDeus:2000cg} 
  J.~Dias de Deus, R.~Ugoccioni and ,
  Phys.\ Lett.\ B {\bf 491}, 253 (2000)
  [hep-ph/0008086].



\bibitem{Kahana:2001ky} 
  D.~E.~Kahana, S.~H.~Kahana and ,
  Phys.\ Rev.\ C {\bf 63}, 031901 (2001).



\bibitem{ALbacete:2010ad} 
  J.~L.~ALbacete, A.~Dumitru and ,
  arXiv:1011.5161 [hep-ph].


\bibitem{Koch:1986ud} 
  P.~Koch, B.~Muller and J.~Rafelski,
  Phys.\ Rept.\  {\bf 142}, 167 (1986).


\bibitem{smes}
M. Ga\'zdzicki, M. I. Gorenstein, Acta Phys. Pol. B. 30, 2705 (1999).

\bibitem{shm}
I. Kuznetsova {\it et al.} J. Phys. G {\bf 35}, 044011 (2008).

\bibitem{stat}
J. Cleymans {\it et al.} Eur. Phys. J. A {\bf 29}, 119
 (2006).

\bibitem{therm}
A. Andronic {\it et al.} Phys. Lett. B {\bf 673}, 142 (2009).

\bibitem{kin}
B. Tomasik {\it et al.} Eur. Phys. J. C {\bf 49}, 115 (2007).

\bibitem{hrghag}
S. Chatterjee {\it et al.} Phys. Rev. C {\bf 81}, 044907
 (2010). 


\bibitem{Shen:2012vn} 
  C.~Shen and U.~Heinz,
  Phys.\ Rev.\ C {\bf 85}, 054902 (2012)
  [Erratum-ibid.\ C {\bf 86}, 049903 (2012)]
  [arXiv:1202.6620 [nucl-th]].


\bibitem{Roy:2012jb} 
  V.~Roy, A.~K.~Chaudhuri and B.~Mohanty,
  Phys.\ Rev.\ C {\bf 86}, 014902 (2012)
  [arXiv:1204.2347 [nucl-th]].


\bibitem{Roy:2012pn} 
  V.~Roy, B.~Mohanty and A.~K.~Chaudhuri,
  arXiv:1210.1700 [nucl-th].


\bibitem{Xu:2011fi} 
  J.~Xu and C.~M.~Ko,
  Phys.\ Rev.\ C {\bf 83}, 034904 (2011)
  [arXiv:1101.2231 [nucl-th]].


\bibitem{Chojnacki:2011hb} 
  M.~Chojnacki, A.~Kisiel, W.~Florkowski and W.~Broniowski,
  Comput.\ Phys.\ Commun.\  {\bf 183}, 746 (2012)
  [arXiv:1102.0273 [nucl-th]].


\bibitem{Kisiel:2005hn} 
  A.~Kisiel, T.~Taluc, W.~Broniowski and W.~Florkowski,
  Comput.\ Phys.\ Commun.\  {\bf 174}, 669 (2006)
  [nucl-th/0504047].




\bibitem{Bass:1998ca} 
  S.~A.~Bass, M.~Belkacem, M.~Bleicher, M.~Brandstetter, L.~Bravina, C.~Ernst, L.~Gerland and M.~Hofmann {\it et al.},
  Prog.\ Part.\ Nucl.\ Phys.\  {\bf 41}, 255 (1998)
  [Prog.\ Part.\ Nucl.\ Phys.\  {\bf 41}, 225 (1998)]
  [nucl-th/9803035].

\bibitem{Bleicher:1999xi} 
  M.~Bleicher, E.~Zabrodin, C.~Spieles, S.~A.~Bass, C.~Ernst, S.~Soff, L.~Bravina and M.~Belkacem {\it et al.},
  J.\ Phys.\ G {\bf 25}, 1859 (1999)
  [hep-ph/9909407].




\bibitem{Adare:2008cg} 
  A.~Adare {\it et al.}  [PHENIX Collaboration],
  Phys.\ Rev.\ C {\bf 77}, 064907 (2008)
  [arXiv:0801.1665 [nucl-ex]].


\bibitem{Abelev:2007ra} 
  B.~I.~Abelev {\it et al.}  [STAR Collaboration],
  Phys.\ Lett.\ B {\bf 655}, 104 (2007)
  [nucl-ex/0703040].


\bibitem{Majumder:2010qh} 
  A.~Majumder and M.~Van Leeuwen,
  Prog.\ Part.\ Nucl.\ Phys.\ A {\bf 66}, 41 (2011)
  [arXiv:1002.2206 [hep-ph]].


\bibitem{Dainese:2004te} 
  A.~Dainese, C.~Loizides and G.~Paic,
  Eur.\ Phys.\ J.\ C {\bf 38}, 461 (2005)
  [hep-ph/0406201].

\bibitem{Vitev:2002pf} 
  I.~Vitev and M.~Gyulassy,
  Phys.\ Rev.\ Lett.\  {\bf 89}, 252301 (2002)
  [hep-ph/0209161].


\bibitem{Armesto:2005iq} 
  N.~Armesto, A.~Dainese, C.~A.~Salgado and U.~A.~Wiedemann,
  Phys.\ Rev.\ D {\bf 71}, 054027 (2005)
  [hep-ph/0501225].


\bibitem{Renk:2011gj} 
  T.~Renk, H.~Holopainen, R.~Paatelainen and K.~J.~Eskola,
  Phys.\ Rev.\ C {\bf 84}, 014906 (2011)
  [arXiv:1103.5308 [hep-ph]].

\bibitem{Wicks:2005gt} 
  S.~Wicks, W.~Horowitz, M.~Djordjevic and M.~Gyulassy,
  Nucl.\ Phys.\ A {\bf 784}, 426 (2007)
  [nucl-th/0512076].

\bibitem{Zhang:2007ja} 
  H.~Zhang, J.~F.~Owens, E.~Wang and X.~-N.~Wang,
  Phys.\ Rev.\ Lett.\  {\bf 98}, 212301 (2007)
  [nucl-th/0701045].

\bibitem{Chatrchyan:2012np} 
  S.~Chatrchyan {\it et al.}  [CMS Collaboration],
  JHEP {\bf 1205}, 063 (2012)
  [arXiv:1201.5069 [nucl-ex]].


\bibitem{ALICE:2013xna} 
  E.~Abbas {\it et al.}  [ALICE Collaboration],
  arXiv:1303.5880 [nucl-ex].

\bibitem{Adamczyk:2012pw} 
  L.~Adamczyk {\it et al.}  [STAR Collaboration],
  arXiv:1212.3304 [nucl-ex].




\bibitem{Adamczyk:2013jei} 
  L.~Adamczyk {\it et al.}  [STAR Collaboration],
  arXiv:1302.6184 [nucl-ex].

\bibitem{CMS:2012qk} 
  S.~Chatrchyan {\it et al.}  [CMS Collaboration],
  Phys.\ Lett.\ B {\bf 718}, 795 (2013)
  [arXiv:1210.5482 [nucl-ex]].


\bibitem{:2013waa} 
L.~Adamczyk, {\it et al.}  [STAR Collaboration],
  arXiv:1301.2187 [nucl-ex].


\bibitem{Aad:2013fja} 
  G.~Aad {\it et al.}  [ATLAS Collaboration],
  arXiv:1303.2084 [hep-ex].






\end{thebibliography}
\end{document}